\documentclass[useAMS,usenatbib]{mn2e}
\usepackage{graphicx}
\usepackage{amssymb, mathtools, epstopdf, url, color}

\newcommand{\rme}{\,\mathrm{e}}
\newcommand{\rmd}{\,\mathrm{d}}
\newcommand{\rmi}{\,\mathrm{i}}
\newcommand{\del}{\partial}

\title[Magnetic damping of stellar oscillations]{Torsional Alfv\'{e}n resonances as an efficient damping mechanism for non-radial oscillations in red giant stars}

\author[S. T. Loi and J. C. B. Papaloizou]{Shyeh Tjing Loi\thanks{E-mail: stl36@cam.ac.uk} and John C. B. Papaloizou\thanks{E-mail: jcbp2@damtp.cam.ac.uk} \vspace{0.2cm} \\
Department of Applied Mathematics and Theoretical Physics, University of Cambridge, Centre for Mathematical Sciences, \\Wilberforce Road, Cambridge CB3 0WA, UK}

\date{Last compiled: \today}

\begin{document}
\maketitle

\begin{abstract}
Stars are self-gravitating fluids in which pressure, buoyancy, rotation and magnetic fields provide the restoring forces for global modes of oscillation. Pressure and buoyancy energetically dominate, while rotation and magnetism are generally assumed to be weak perturbations and often ignored. However, observations of anomalously weak dipole mode amplitudes in red giant stars suggest that a substantial fraction of these are subject to an additional source of damping localised to their core region, with indirect evidence pointing to the role of a deeply buried magnetic field. It is also known that in many instances the gravity-mode character of affected modes is preserved, but so far no effective damping mechanism has been proposed that accommodates this aspect. Here we present such a mechanism, which damps the oscillations of stars harbouring magnetised cores via resonant interactions with standing Alfv\'{e}n modes of high harmonic index. The damping rates produced by this mechanism are quantitatively on par with those associated with turbulent convection, and in the range required to explain observations, for realistic stellar models and magnetic field strengths. Our results suggest that magnetic fields can provide an efficient means of damping stellar oscillations without needing to disrupt the internal structure of the modes, and lay the groundwork for an extension of the theory of global stellar oscillations that incorporates these effects.
\end{abstract}


\section{Introduction}\label{sec:intro}
Surface convection in many stars stochastically excites global oscillations (normal modes), which can be detected through the intensity fluctuations associated with temperature variations induced at the stellar surface \citep{Houdek2015}. These normal modes can be regarded as standing superpositions of waves associated with restoring forces produced by pressure, buoyancy, the Coriolis force (in the presence of rotation) and the Lorentz force (if the star harbours a magnetic field). Pressure and buoyancy effects dominate energetically over those produced by rotation and magnetic fields, and so to a first approximation one identifies in the asymptotic limit of high and low frequencies two types of modes: p-modes, restored mainly by pressure and associated with large surface displacements; and g-modes, restored mainly by buoyancy and associated with large interior displacements \citep{Deubner1984}. In reality, and particularly in the case of evolved stars, modes are not purely one type or the other but have mixed character, exhibiting large fluid displacements both near the surface and in the deep interior \citep{Osaki1975}.

The fluid displacement field $\boldsymbol{\xi}(\mathbf{r},t)$ at the point with position vector $\mathbf{r}$ associated with a normal mode of oscillation can be described by a spatial amplitude function modulated by a time-harmonic component $\exp(-\rmi \omega t)$. If rotation and magnetic fields are weak, which is the case for the vast majority of stars, then there is negligible departure of the stellar background from spherical symmetry. This allows one to expand the spatial part in terms of vectorial spherical harmonics, i.e.~the overall fluid displacement can be written
\begin{align}
  \boldsymbol{\xi}(\mathbf{r}, t) = \left[ \xi_r Y_\ell^m \hat{\mathbf{r}} + \xi_h \nabla Y_\ell^m + \xi_T \hat{\mathbf{r}} \times \nabla Y_\ell^m \right] \exp(-\rmi \omega t) \:, \label{eq:xi_VSH}
\end{align}
where we adopt spherical polar coordinates $(r, \theta, \phi)$ and there is an implicit summation over spherical harmonics $Y_\ell^m(\theta,\phi)$. Radial dependencies are captured solely by the scalar functions $\xi_r(r)$, $\xi_h(r)$ and $\xi_T(r)$, which describe displacements in three mutually orthogonal directions for given $\ell$ and $m$. The first two terms on the RHS of Eq.~(\ref{eq:xi_VSH}) are collectively referred to as the spheroidal component, while the third (involving $\xi_T$) is the torsional component. Mathematically, spheroidal motions are those for which $(\nabla \times \boldsymbol{\xi})_r = 0$, the subscript $r$ denoting the radial component. Physically, these are motions that involve deformation but no twist. Torsional motions have $\nabla \cdot \boldsymbol{\xi} = 0$, and correspond to motions that involve twist but no deformation. 

In the absence of rotation and magnetic fields, one can show from the fluid equations of motion that for $\omega \neq 0$, $\xi_T = 0$, implying that pressure and buoyancy are only capable of restoring spheroidal motions. However, it is possible to access the third spatial degree of freedom (torsional motions) in the presence of rotation and/or magnetic fields. In this work we ignore rotation. Our aim is to investigate the dynamical consequences of interactions between torsional motions restored by the Lorentz force with the usual spheroidal (i.e., p- and g-) modes. We deal with the limit where the magnetic field is sufficiently weak that it does not disrupt the structure of the spheroidal modes. We find that resonant interactions between the two types of modes can provide an efficient energy sink for spheroidal motions. This source of damping may be potentially important for explaining the anomalously low amplitudes of non-radial (particularly dipole) modes observed in some evolved stars.

The existence of red giant stars exhibiting low amplitudes of their dipole ($\ell = 1$) modes was first reported by \citet{Mosser2012_mn2e}, accounting for roughly 20\% of their sample. For the remainder, the higher amplitudes of their $\ell = 1$ modes are consistent with the primary source of damping being convection alone. Given that the red giant population appears to be divided into those with either high or low $\ell = 1$ mode amplitudes, with relatively few intermediate cases, we shall refer to this as the \textit{dipole dichotomy} problem. While the frequencies of the low-amplitude $\ell = 1$ modes are close to those predicted by the usual asymptotic relation \citep{Tassoul1980, Gough1986} obeyed by the remainder of the sample \citep{Mosser2011_mn2e, Mosser2012b_mn2e}, their widths are considerably larger \citep{Garcia2014_mn2e}, suggesting that the $\ell = 1$ modes of these stars are subject to an additional source of damping. Follow-up analyses by \citet{Stello2016a} established that a dichotomy also exists for the $\ell = 2$ modes, but to a lesser extent than $\ell = 1$. Radial ($\ell = 0$) modes appear to be unaffected. As argued by \citet{Garcia2014_mn2e}, sources of damping such as turbulent viscosity localised to the convective envelope should affect all low-degree modes to a similar extent, and so to selectively affect non-radial modes the extra source of damping needs to be localised to the core. A further piece of evidence is that the behaviour is mass-dependent \citep{Mosser2012_mn2e}, being restricted to stars more massive than 1.1\,M$_\odot$ \citep{Stello2016_mn2e}. This is roughly the threshold mass above which stars on the main sequence possess convective rather than radiative cores.

Convective regions of stars are strongly associated with dynamo action and therefore the existence of a magnetic field \citep{Proctor1994, Charbonneau2001}. Numerical simulations suggest that convective core dynamos in massive stars may generate field strengths of 10--100\,kG or more \citep{Brun2005, Featherstone2009}. The long timescales of magnetic diffusion in stellar cores, which greatly exceed nuclear timescales, suggest that after the dynamo ceases at the end of the star's main sequence life the field should relax into a long-lived equilibrium state if sufficiently large-scale magnetic structure can be retained during the cessation phase. Although there have been many previous works investigating the effects of magnetic fields on stellar oscillations \citep[e.g.,][]{Campbell1986, Cunha2000, Rincon2003, Reese2004, Lee2007}, these are not directly applicable here as they have mainly been concerned with cases where the magnetic field of interest extends beyond the star and is dynamically significant only in a thin layer near the surface. Prior to the discovery of the dipole dichotomy problem, the existence of core-confined fields, though not generally disputed, was not considered to give rise to observable consequences. Until recently, very little attention has been paid to their possible influence on stellar oscillations.

The link to main-sequence dynamo action led to suggestions that the mechanism behind the dipole dichotomy might involve a deeply buried magnetic field \citep{Garcia2014_mn2e}. Follow-up theoretical work by \citet{Fuller2015} and \citet{Lecoanet2016} has established that if the magnetic field strength exceeds a critical threshold, complete conversion of gravity waves to magnetoacoustic waves occurs, which then dissipate within the core (damping processes associated with conversion between different wave modes have been previously been investigated mainly in the context of the solar atmosphere, e.g.~\citet{Spruit1992}). This acts to selectively damp non-radial modes, while also implying that modes having the character of g-modes could not be constructed (only pure p-modes could exist). An additional prediction is that affected p-modes should be magnetically split to an extent comparable to g-mode period spacings and rotational splittings \citep{Cantiello2016}. However, more detailed analyses of the observational data performed by \citet{Mosser2016} indicate that (i) additional splitting of this sort is not seen, (ii) the measured mode amplitudes are inconsistent with total energy conversion, and (iii) in many instances the mixed character of the modes is retained. The current consensus is that an independent mechanism is required to explain the existence of stars possessing mixed modes with weak amplitudes, and where the amplitude depression is only partial.

In this work we present a new mechanism for damping spheroidal modes involving resonant interactions with torsional Alfv\'{e}n modes localised to the magnetised core (a similar idea was briefly speculated on by \citet{Reese2004} in the context of roAp stars, but this has not been pursued further in any context). No critical field strength is necessary: our mechanism is capable of operating in the regime below the threshold required by \citet{Fuller2015}. Moreover, no disruption to the structure of the spheroidal modes is required, implying that the mixed character of modes can be retained. Damping rates are a function of several parameters including the field strength, so in general one expects only partial energy loss. In Section \ref{sec:models} we describe the background stellar model and magnetic field configuration used. In Section \ref{sec:mechanism} we explain the details of the damping mechanism, and present quantitative results for the application of this to a 2\,M$_\odot$ red giant model in Section \ref{sec:results}. We discuss observational consequences and limitations in Section \ref{sec:discuss}. We conclude in Section \ref{sec:conclude}.

\section{Models}\label{sec:models}
\subsection{Red giant stellar model}
\begin{figure*}
  \centering
  \includegraphics[clip=true, trim=1cm 0cm 1cm 0cm, width=\textwidth]{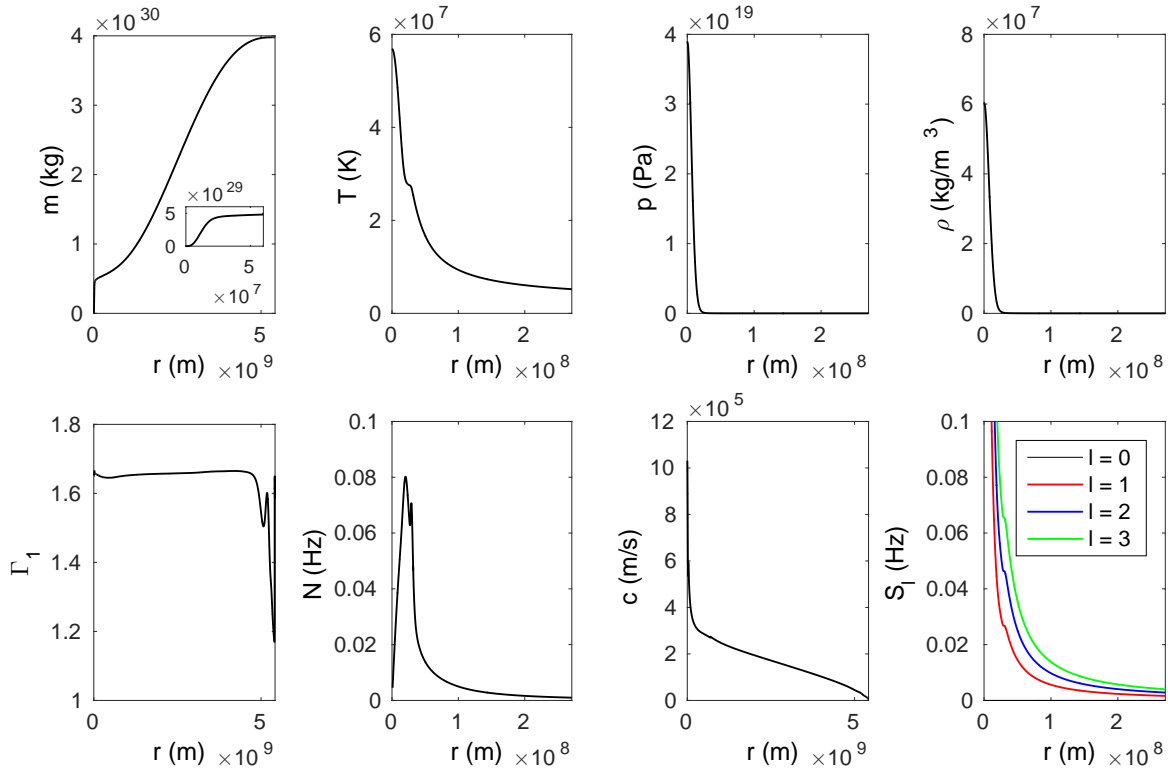}
  \caption{From left to right, top to bottom: plots of the enclosed mass, temperature, pressure, density, adiabatic index, buoyancy frequency, sound speed and Lamb frequency profiles, for a 2$M_\odot$ red giant stellar model (generated by CESAM). Note that the temperature, pressure, density and buoyancy/Lamb frequency plots are only shown up to 5\% of the stellar radius. In the plot of enclosed mass (top left), an inset plot zooming in to 1\% of the stellar radius has been included to illustrate the high central mass concentration.}
  \label{fig:profiles}
\end{figure*}

We will illustrate our mechanism in the context of a 2\,M$_\odot$ red giant model whose background profiles were generated by the CESAM (Code d'Evolution Stellaire Adaptatif et Modulaire) stellar evolutionary code \citep{Morel1997}. We obtained the parameter grids from an online source\footnote{\url{https://www.astro.up.pt/helas/stars/cesam/A/data/}}. The various background quantities are shown plotted in Fig.~\ref{fig:profiles}. The age of the model is 963\,Myr, when the star is at an intermediate position in its ascent along the red giant branch (RGB). At this stage the radius of the star is 7.7\,R$_\odot$ and the dynamical timescale (given by $\sqrt{R_*^3/GM_*}$, where $R_*$ and $M_*$ are the stellar radius and mass) is 6.8\,hr. Although our proposed mechanism is quite general, we have chosen to demonstrate it using a reasonably realistic stellar model to obtain meaningful estimates of the damping rates for comparison with observation.

\subsection{Magnetic field configuration}
Following cessation of convective fluid motions and therefore the dynamo at the end of the main sequence, the magnetic field is expected to relax rapidly into an equilibrium state (note that the Lorentz force during dynamo operation is strong enough to influence the velocity field) \citep{Braithwaite2015}. The timescale of relaxation is the Alfv\'{e}n travel time across the core, which can be as short as 1\,yr (for a core diameter of $R_c \sim 100$\,Mm, $B \sim 10$\,kG and $\rho \sim $100\,g\,cm$^{-3}$). Where they exist in nature, fields in non-convective regions should therefore obey the force-balance condition
\begin{align}
  \nabla p + \rho \nabla \Phi = \frac{1}{\mu_0} (\nabla \times \mathbf{B}) \times \mathbf{B} \:, \label{eq:MHS_cond}
\end{align}
where $p$, $\rho$, $\Phi$ and $\mathbf{B}$ are the pressure, density, gravitational potential and magnetic field, respectively. If that were not the case then they would evolve towards such a state on the Alfv\'{e}n timescale.

To model a physically realistic equilibrium field that might be found in a red giant core, one seeks a solution to Eq.~(\ref{eq:MHS_cond}) that (i) is spatially confined, (ii) is finite throughout, (iii) is continuous at the boundary interior to which the field is confined, and (iv) is stable. The third condition is needed to avoid infinite current sheets on the boundary. It has been shown that purely poloidal and purely toroidal fields are unstable \citep{Tayler1973, Markey1973, Flowers1977}, and so for stability, magnetic equilibria necessarily involve a mixture of poloidal and toroidal components. Numerical studies, which find that random initial fields tend to settle into mixed poloidal-toroidal configurations with roughly comparable strengths of the two components, support this notion \citep{Braithwaite2004, Braithwaite2006}. Notably, this means that a simple dipole field is inadequate for our purposes, since it violates all four criteria. More generally, it can be shown that magnetic fields which are force free throughout, spatially confined and continuous on the boundary must vanish identically \citep{Roberts1967, Braithwaite2015}, and so we require a non-force free configuration.

Early analytic work by \citet{Prendergast1956} successfully obtained an axisymmetric, mixed poloidal-toroidal solution satisfying the first three criteria. Though originally derived for incompressible stars, its extension to compressibility produces a qualitatively similar result \citep{Braithwaite2006, Duez2010a}. Although we do not check for stability of this configuration for the red giant model considered here, its stability for $n = 3$ polytropes has previously been verified numerically \citep{Duez2010}. In addition we expect that confined fields with poloidal and toroidal components of comparable strength should in general be adequately characterised by the configuration we employ. To obtain this field solution, one begins by introducing the poloidal flux function $\psi(r,\theta)$, in terms of which an axisymmetric magnetic field $\mathbf{B} = (B_R, B_\phi, B_z)$ may be written
\begin{align}
  \mathbf{B} = \frac{1}{R} \nabla \psi \times \hat{\boldsymbol{\phi}} + B_\phi \hat{\boldsymbol{\phi}} \:. \label{eq:B_psi}
\end{align} 
Note that $(R,\phi,z)$ will be used to refer to cylindrical polar coordinates, while $(r,\theta,\phi)$ are spherical polar coordinates (the $\phi$ coordinate is the same for each and refers to the azimuthal direction). Physically, $\psi$ is a flux surface label, meaning that poloidal projections of the field lines are the level surfaces of $\psi$. Substituting Eq.~(\ref{eq:B_psi}) into Eq.~(\ref{eq:MHS_cond}) and applying the appropriate vector identities, one can show that the quantity $F \equiv RB_\phi$ is invariant on flux surfaces, i.e.~$F = F(\psi)$. If in addition we assume a barotropic configuration, we arrive at a nonlinear PDE known as the Grad-Shafranov equation:
\begin{align}
  \Delta^* \psi + F \frac{\rmd F}{\rmd \psi} = -\mu_0 \rho R^2 G \:, \label{eq:GSE} \\
  \shortintertext{where}
  \Delta^* \equiv \frac{\del^2}{\del R^2} + \frac{\del^2}{\del z^2} - \frac{1}{R} \frac{\del}{\del R} \nonumber \\
  \equiv \frac{\partial^2}{\partial r^2} + (1-\mu^2) \frac{\partial^2}{\partial \mu^2} \:, \label{eq:GSE_op}
\end{align}
$\mu \equiv \cos \theta$, and $G = G(\psi)$ is another flux surface invariant. Neither the functional forms of $F$ or $G$ are constrained within ideal MHD (different choices correspond to different equilibria), and so following \citet{Prendergast1956}, a simplifying choice is to set $F(\psi) = \lambda \psi$ and $G(\psi) = -\beta/\mu_0$, where $\lambda$ and $\beta$ are constants. Separating $\psi(r,\theta) = \Psi(r) \sin^2 \theta$ turns Eq.~(\ref{eq:GSE}) into
\begin{align}
  \Psi'' - \left( \frac{2}{r^2} - \lambda^2 \right) \Psi = \beta \rho r^2 \:, \label{eq:Bessel}
\end{align}
which is an inhomogeneous, second-order ODE that can be solved using the method of Green's functions. Applying the boundary conditions $\Psi(0) = 0$, $\Psi(r_1) = 0$ and $\Psi'(r_1) = 0$, where $r_1$ is the boundary of the field region, this produces the result
\begin{align}
  \Psi(r) = \frac{\beta \lambda r}{j_1(\lambda r_1)} \left[ f(r, r_1; \lambda) \int_0^r \rho(\xi) \xi^3 j_1(\lambda \xi) \rmd \xi \right. \nonumber \\
    \qquad + \left. j_1(\lambda r) \int_r^{r_1} \rho(\xi) \xi^3 f(\xi, r_1; \lambda) \rmd \xi \right] \:, \label{eq:Prendergast} \\
  \shortintertext{where}
  f(\xi_1, \xi_2; \lambda) \equiv j_1(\lambda \xi_2) y_1(\lambda \xi_1) - j_1(\lambda \xi_1) y_1(\lambda \xi_2) \:, \label{eq:aux_f}
\end{align}
$j_1$ and $y_1$ are spherical Bessel functions of the first and second kind, and $\lambda$ is a root of
\begin{align}
  \int_0^{r_1} \rho(\xi) \xi^3 j_1(\lambda \xi) \rmd \xi = 0 \label{eq:lambda_cond} \:.
\end{align}
The oscillatory nature of $j_1$ means that more than one possible value of $\lambda$ may satisfy Eq.~(\ref{eq:lambda_cond}); we chose to use the smallest one. We have set $r_1 = 0.005 R_*$ (the inferred boundary of what used to be the convective core) on the basis of inspection of the stellar profiles. This corresponds to a mass coordinate of 0.11\,$M_*$. For the $\rho(r)$ profile of the red giant, we obtain $\lambda = 2643.2 R_*^{-1}$, yielding comparable poloidal and toroidal field strengths (the maximum values of each of these differ by less than a per cent). 

\begin{figure}
  \centering  
  \includegraphics[width=0.8\columnwidth]{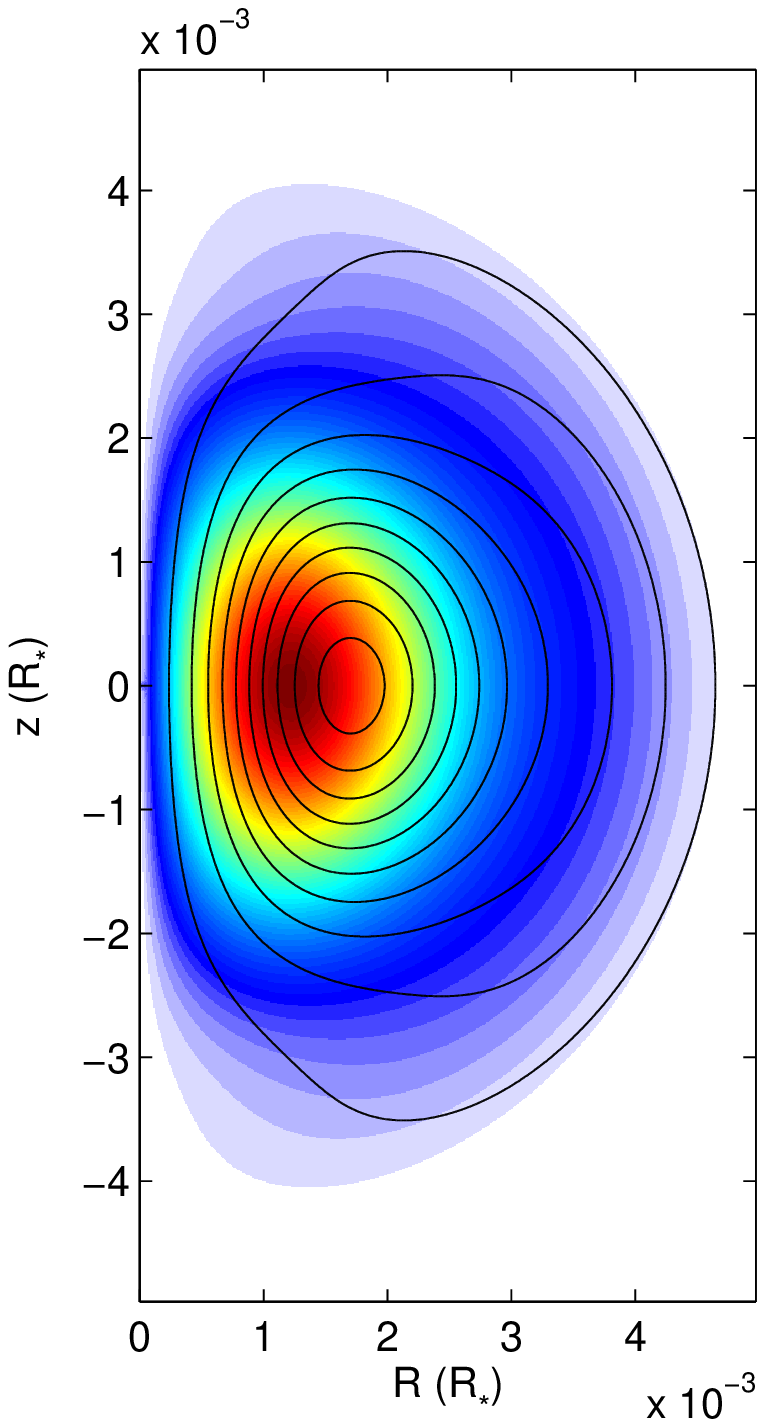}
  \includegraphics[width=0.8\columnwidth]{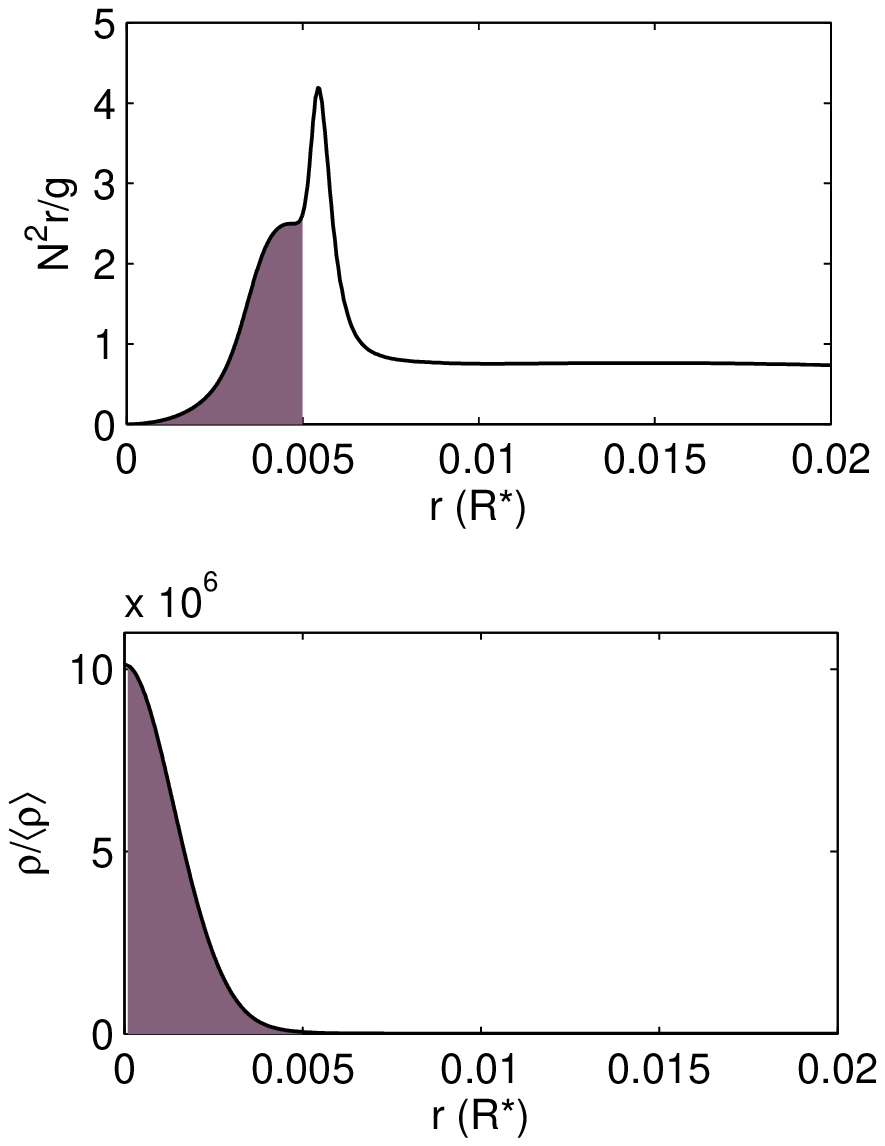}
  \caption{The Prendergast magnetic field solution (top) calculated over the assumed core region (0.005 of the stellar radius). This region is shown shaded in the bottom two panels, which plot the dimensionless squared buoyancy frequency and mass density profiles near the centre of the star. In the top panel, a selection of magnetic flux surfaces (poloidal field loops) are shown as black lines, while the underlying colour represents the strength of the toroidal component. The absolute scaling of the field at this stage is arbitrary; illustrated here is just the overall geometry. Note that the solution is axisymmetric and so only a meridional half-plane needs to be shown.}
  \label{fig:Prendergast}
\end{figure}

The components of $\mathbf{B}$ in terms of $\Psi$ are given in spherical polar coordinates by
\begin{align}
  B_r(r,\theta) &= \frac{2}{r^2} \Psi(r) \cos \theta \:, \nonumber \\
  B_\theta(r,\theta) &= -\frac{1}{r} \Psi'(r) \sin \theta \:, \label{eq:field_compts} \\
  B_\phi(r,\theta) &= -\frac{\lambda}{r} \Psi(r) \sin \theta \:. \nonumber
\end{align}
The corresponding field configuration, which we will refer to as \textit{Prendergast's solution}, is displayed in Fig.~\ref{fig:Prendergast}.

Prendergast's solution qualitatively resembles a dipole in its angular dependence, but unlike a dipole field possesses no singularity, has a mixed poloidal-toroidal topology, and vanishes smoothly at the spherical boundary $r = r_1$ in all three components of $\mathbf{B}$. Beyond this radius we set $\mathbf{B} = \mathbf{0}$, i.e.~we neglect the envelope field under the assumption that this is much weaker than the core field. The overall strength of the Prendergast field is controlled through the parameter $\beta$, which sets the amplitude of the field but not its shape. Inspection of the CESAM grid of models for this star shows that the core contracts by roughly a factor of 10 in radius from 403 Myr (on the main sequence) to 963 Myr. Under conservation of magnetic flux, central field strengths should increase by a factor of about 100. If magnetic field strengths on the main sequence were 10--100 kG, then one expects field strengths in the red giant core to be of order several MG. We have set $\beta$ such that the central field strength is 4 MG.

\section{Alfv\'{e}n Resonance Damping Mechanism}\label{sec:mechanism}
In this section we present a mechanism for damping spheroidal modes through interaction with an embedded magnetic field, illustrating this for a red giant containing a Prendergast field in the core. This section is structured as follows. First, we isolate eigenmodes of oscillation that in the limit of a weak magnetic field are purely torsional in nature \citep[e.g.][]{Mestel2012} and correspond to standing Alfv\'{e}n waves localised to the field region (Section \ref{sec:torsional}). We show that these couple to the spheroidal modes through the Lorentz force (Section \ref{sec:coupling}). We then incorporate viscous and Ohmic dissipation and show that this produces a damping of the torsional modes, implying that the torsional problem is one of a driven-damped mechanical oscillator (Section \ref{sec:dissipation}). Under resonant conditions, the rates of driving and dissipation for such a system exactly balance. In the context of the stellar problem, this means that where resonances between spheroidal and torsional modes exist, the energy dissipated equals the work done against the Lorentz force by the spheroidal motions. Integrating this over the star, we arrive at an analytical expression for the overall damping rate $\gamma$ of a spheroidal mode (Section \ref{sec:overall_damping}). Throughout this work we assume linearity of the fluid motions, and specialise to the case of axisymmetric modes.

\subsection{Torsional Alfv\'{e}nic oscillations}\label{sec:torsional}
The equation of motion of a driven oscillator can be written
\begin{align}
  \frac{\del^2 \boldsymbol{\xi}}{\partial t^2} + \mathcal{L}[\boldsymbol{\xi}] = \mathbf{S}(\mathbf{r},t) \:, \label{eq:driven_osc}
\end{align}
where $\boldsymbol{\xi}$ denotes mechanical displacement, $\mathcal{L}$ is a linear operator containing the spatial derivatives of $\boldsymbol{\xi}$ and the source term $\mathbf{S}$ represents the external driving/forcing. Here $\mathbf{S}$ is regarded as being external because it does not depend on $\boldsymbol{\xi}$, although it may be a function of position $\mathbf{r}$ and time $t$. The normal modes of the oscillator are the solutions of Eq.~(\ref{eq:driven_osc}) with $\mathbf{S} = \mathbf{0}$ (the homogeneous problem). Imposing a time-harmonic dependence $\boldsymbol{\xi} \propto \exp(-\rmi \omega t)$, this corresponds to the eigenproblem $\mathcal{L}[\boldsymbol{\xi}] = \omega^2 \boldsymbol{\xi}$, satisfied for only special values of $\omega^2 = \omega_0^2$. These are the natural frequencies of the oscillator, and the associated forms of $\boldsymbol{\xi}$ are the eigenfunctions of the system.

The fluid equation of motion in the absence of rotation can be written
\begin{align}
  \rho \left( \frac{\del \mathbf{u}}{\del t} + \mathbf{u} \cdot \nabla \mathbf{u} \right) = -\nabla p - \rho \nabla \Phi - \frac{1}{2} \nabla B^2 + (\mathbf{B} \cdot \nabla) \mathbf{B} \:, \label{eq:EoM}
\end{align}
where $\mathbf{u} = D\boldsymbol{\xi} / Dt$ is the fluid velocity and we have absorbed the usual $\mu_0$ factors into the definition of $\mathbf{B}$. The last two terms on the RHS of Eq.~(\ref{eq:EoM}) correspond to the magnetic pressure and tension, respectively. We here adopt the Cowling approximation under which the gravitational potential is fixed and depends only on $r$. Then upon linearising and taking the curl of Eq.~(\ref{eq:EoM}) the $r$-components of the first three terms on the RHS vanish, leaving magnetic tension as the only force capable of restoring torsional motions. In the axisymmetric case, which we focus on here, the torsional direction corresponds to the $\phi$ (azimuthal) direction (we comment on the non-axisymmetric case in Section \ref{sec:limits}). Consider now the torsional component of Eq.~(\ref{eq:EoM}), which linearises to give
\begin{align}
  \rho_0 \frac{\del^2 \xi_\phi}{\del t^2} = \frac{\mathbf{B}_0}{R} \cdot \nabla (RB_\phi') + \frac{\mathbf{B}'}{R} \cdot \nabla (RB_{0\phi}) \:. \label{eq:EoM_lin_tor}
\end{align}
Subscript 0's denote static background quantities, while primes denote (small) time-dependent perturbations about the background. Using the linearised induction equation $\mathbf{B}' = \nabla \times (\boldsymbol{\xi} \times \mathbf{B}_0)$, this allows us to express Eq.~(\ref{eq:EoM_lin_tor}) in terms of just $\boldsymbol{\xi}$ and background quantities. This can be written in the form
\begin{align}
  \frac{\del^2 \xi_\phi}{\del t^2} &+ \mathcal{L}_T[\xi_\phi] = \frac{f_\text{TS}}{\rho_0} \:, \label{eq:EoM_tor_osc} \\
  \shortintertext{where}
  \mathcal{L}_T[\xi_\phi] &= -\frac{\mathbf{B}_0}{\rho_0 R} \cdot \nabla \left[ R^2 \mathbf{B}_0 \cdot \nabla \left( \frac{\xi_\phi}{R} \right) \right] \:, \label{eq:tor_op} \\
  f_\text{TS} &= -\frac{\mathbf{B}_0}{R} \cdot \nabla \left[ R^2 \boldsymbol{\xi} \cdot \left( \frac{B_{0\phi}}{R} \right) + RB_{0\phi} (\nabla \cdot \boldsymbol{\xi}) \right] \nonumber \\
  &\qquad + \frac{1}{R} \left[ \nabla \times (\boldsymbol{\xi} \times \mathbf{B}_0) \right] \cdot \nabla (RB_{0\phi}) \:. \label{eq:f_TS}
\end{align}
See that $f_\text{TS}$ depends only on the spheroidal displacement $\boldsymbol{\xi}_S \equiv (\xi_R, 0, \xi_z)$, not $\xi_\phi$, and can thus be regarded as a forcing term in Eq.~(\ref{eq:EoM_tor_osc}) provided that the spheroidal displacement is assumed known, which will effectively be the case when the magnetic field is weak. In that case the spheroidal modes will be relatively unperturbed.

\begin{figure}
  \centering
  \includegraphics[width=0.7\columnwidth]{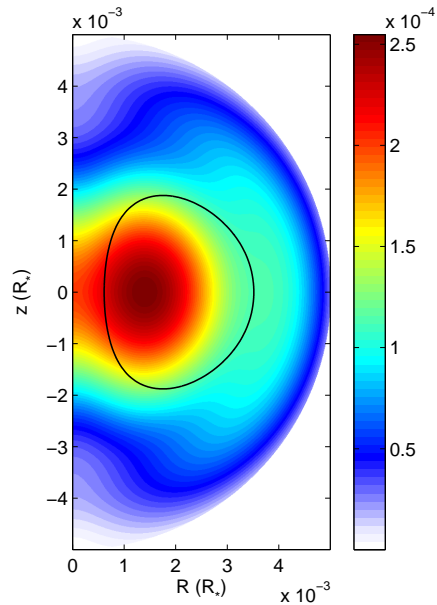}
  \includegraphics[width=\columnwidth]{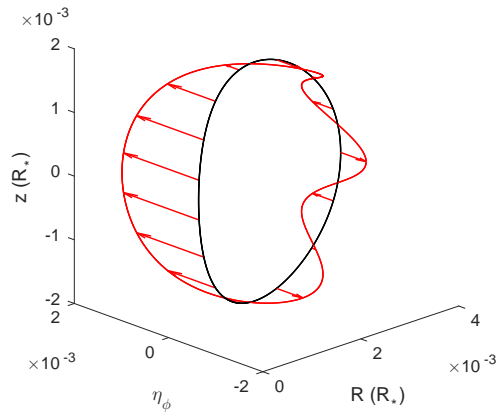}
  \caption{Spatial distribution of the Alfv\'{e}n speed overlaid with an arbitrary field loop (top), and the amplitude function for the $j = 7$ eigenmode on that loop (bottom). Colour bar units are in terms of the dynamical speed $\sqrt{GM_*/R_*}$. In the bottom panel, the equilibrium position of the field line is shown in black and the displaced position in red. Arrows are an aid to visualising the direction of the displacement.}
  \label{fig:vAmag_X150}
\end{figure}

Let us examine the operator $\mathcal{L}_T$ more closely. Although at first glance this appears to depend on three spatial dimensions, notice that $\mathbf{B}_0 \cdot \nabla = B_p \del/\del s$, where $B_p$ is the magnitude of the poloidal component of $\mathbf{B}$, and $s$ is arc length (i.e.~physical distance) along the poloidal projections of the field lines. Hence the problem is intrinsically one-dimensional. Rescaling to a new distance coordinate $\sigma$ obeying $\rmd \sigma / \rmd s = 1 / (R^2 B_p)$, the eigenproblem reduces to
\begin{align}
  \frac{\del^2 \eta_\phi}{\del t^2} = v_A^2 \frac{\del^2 \eta_\phi}{\del \sigma^2} \:, \label{eq:1DWE}
\end{align}
where $\eta_\phi \equiv \xi_\phi/R$ is a new scaled fluid displacement and $v_A^2 \equiv 1/(\rho_0 R^4)$. One recognises Eq.~(\ref{eq:1DWE}) as the one-dimensional wave equation with spatially varying advection speed $v_A = v_A(\sigma)$, which can be identified as the Alfv\'{e}n speed with respect to the new coordinates. From here it becomes more convenient to work in terms of $\eta_\phi$ rather than $\xi_\phi$. The particular form of Eq.~(\ref{eq:1DWE}) allows the solutions to be understood intuitively as standing waves on stretched 1D loops. These are quantised vibrations whose frequencies increase as the spatial scale decreases.

We solved Eq.~(\ref{eq:1DWE}) as a matrix eigenvalue problem on a discrete 1D grid for each flux surface, with periodic boundary conditions and spatial derivatives approximated by centred differences. We calculated the eigenmodes $X_j(\sigma, \psi)$ and eigenfrequencies $\omega_{0,j}^2(\psi)$ on 1000 evenly-spaced (in $\psi$) flux surfaces with 5000 uniformly-spaced (in $\sigma$) points on each surface. Here $j \in \mathbb{Z}^+$ is the harmonic index. Although the total number of eigenfunctions obtainable by this method equals the number of grid points, the accuracy of the solutions is expected to degrade for larger $\omega_{0,j}^2$ where spatial scales of the associated eigenfunctions become too small to be adequately resolved. On each flux surface we restricted the eigenfunctions used in further analysis to the 1000 having the lowest eigenfrequencies.

\begin{figure}
  \centering
  \includegraphics[width=0.85\columnwidth]{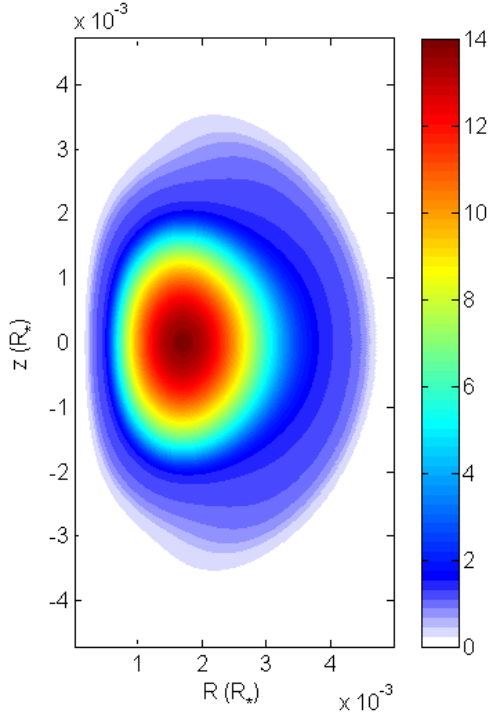}
  \caption{Spatial distribution of the $j = 300$ eigenfrequencies. Colour bar units are in terms of the dynamical frequency $\sqrt{GM_*/R_*^3}$. Since eigenmodes are localised to individual flux surfaces, $\omega_{0,j}$ is constant on any flux surface for given $j$. Smaller flux surfaces tend to have higher $\omega_{0,j}$ for fixed $j$.}
  \label{fig:om_300}
\end{figure}

The parameter $\beta$ was chosen so as to produce the distribution of the Alfv\'{e}n speed, $v_A$, shown in the upper panel of Fig.~\ref{fig:vAmag_X150}. A selected eigenmode is illustrated in the lower panel of Fig.~\ref{fig:vAmag_X150}. Since the problem is axisymmetric, the spatial amplitude function need only be displayed on a poloidal field loop, corresponding to a longitudinal slice of the flux surface. The full solution is obtained by sweeping each loop in a circle about the axis of symmetry (here the $z$-axis). The motion can be envisaged as segments of each flux surface (which are tori in 3D) twisting with respect to others. For a given wave speed $v_A$, one expects $\omega_{0,j}$ for fixed $j$ to increase as the length of the field loop shrinks. This is indeed observed in our model: the spatial distribution of $\omega_{0,j}$ for $j = 300$ is shown in Fig.~\ref{fig:om_300}. 

\begin{figure}
  \centering
  \includegraphics[width=\columnwidth]{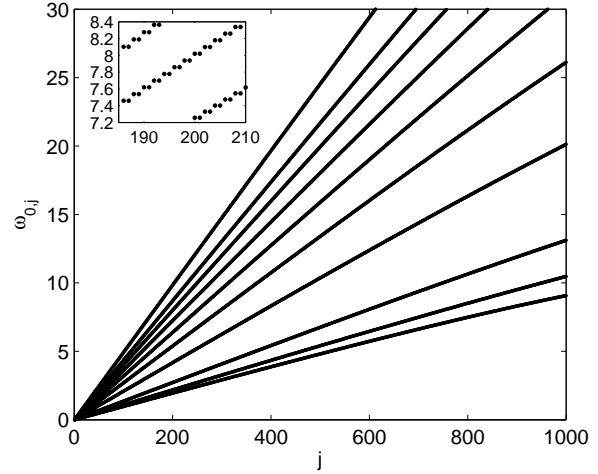}
  \caption{The torsional spectrum calculated for 10 evenly-spaced (in $\psi$) flux surfaces. Each track corresponds to one flux surface, and flux surfaces of lower tracks enclose those of higher ones. Although apparently continuous, the tracks are in fact made up of discrete points, since $j \in \mathbb{Z}^+$. The discreteness can be seen in the inset plot, which zooms in to a small portion of the overall spectrum. The paired structure reflects approximately degenerate modes, which have equal numbers of nodes but are odd and even versions of one another. Frequencies are given in units of the dynamical frequency, $\sqrt{GM_*/R_*^3}$.}
  \label{fig:Teig_10FLs}
\end{figure}

For a given flux surface, one also expects $\omega_{0,j}$ to increase with $j$. Only an even number of nodes is allowed for vibrations on a loop, so $j = 1$ has zero nodes, $j = 2,3$ have two, $j = 4,5$ have four, and so on. Hence $j$ is roughly proportional to the number of wavelengths around the loop, but there is a paired structure to the spectrum. This can be seen in the inset to Fig.~\ref{fig:Teig_10FLs}, which plots $\omega_{0,j}$ versus $j$ for a selection of flux surfaces. Note that despite having equal numbers of nodes, each pair still correspond to distinct eigenmodes, these being odd and even versions of one another (e.g.~for a constant $v_A$ they would be the sine and cosine solutions), with slightly different eigenfrequencies. Note that similar behaviour of the periodic solutions of the Mathieu equation occurs. The overall slope of $\omega_{0,j}$ versus $j$ should also be proportional to the fundamental frequency (larger for smaller loops, in line with the picture of a vibrating string). As can be seen in Fig.~\ref{fig:Teig_10FLs}, this is indeed the case. 

\subsection{Coupling with spheroidal motions}\label{sec:coupling}
The perturbation to the Lorentz force can be subdivided into terms that depend on the spheroidal displacement $\boldsymbol{\xi}_S$ but not the torsional displacement $\xi_\phi$, and the terms that depend on $\xi_\phi$ but not $\boldsymbol{\xi}_S$ (this is possible because only terms linear in $\boldsymbol{\xi}$ are retained). Writing this out in components, we can express this separation as
\begin{align}
  \mathbf{f}_\text{S}(\boldsymbol{\xi}) &= \mathbf{f}_\text{SS}(\boldsymbol{\xi}_S) + \mathbf{f}_\text{ST}(\xi_\phi) \:, \nonumber \\ 
  f_\text{T}(\boldsymbol{\xi}) &= f_\text{TS}(\boldsymbol{\xi}_S) + f_\text{TT}(\xi_\phi) \:, \label{eq:fS_and_fT}
\end{align}
where $\mathbf{f}_\text{S}$ and $f_\text{T}$ and the spheroidal and torsional components to the Lorentz force perturbation, $\mathbf{f}_\text{SS}$ are the terms in $\mathbf{f}_\text{S}$ that depend only on $\boldsymbol{\xi}_S$, $\mathbf{f}_\text{ST}$ are those that depend only on $\xi_\phi$, etc. This allows us to observe the following coupled structure of the equations of motion:
\begin{align}
  \frac{\del^2 \boldsymbol{\xi}_S}{\del t^2} + \mathcal{L}_S[\boldsymbol{\xi}_S] &= \frac{\mathbf{f}_\text{ST}(\xi_\phi)}{\rho_0} \label{eq:EoM_sph_lin} \\
  \frac{\del^2 \xi_\phi}{\del t^2} + \mathcal{L}_T[\xi_\phi] &= \frac{f_\text{TS}(\boldsymbol{\xi}_S)}{\rho_0} \:, \label{eq:EoM_tor_lin_2} \\
  \shortintertext{where}
  \mathcal{L}_S[\boldsymbol{\xi}_S] &= \frac{1}{\rho_0} \left[ \nabla p' + \rho' \nabla \Phi_0 - \mathbf{f}_\text{SS}(\boldsymbol{\xi}_S) \right] \:, \label{eq:sph_op} \\
  \mathcal{L}_T[\xi_\phi] &= -\frac{f_\text{TT}(\xi_\phi)}{\rho_0} \:. \label{eq:tor_op_2}
\end{align}

We see that the coupling between spheroidal and torsional motions is provided by the Lorentz force. To first order, given that the Lorentz force is much smaller than the forces of pressure and buoyancy, the $\mathbf{f}_\text{SS}$ term in Eq.~(\ref{eq:sph_op}) can be neglected. This is akin to assuming that the magnetic field has negligible effect on the spheroidal eigensolution (i.e.~we still get the usual p- and g-modes). An important term not included explicitly in Eq.~(\ref{eq:EoM_sph_lin}) is the forcing associated with (purely spheroidal) convective motions, the source of energy for the whole system. Given that the coupling from spheroidal motions into torsional motions and back into spheroidal motions is a second-order process, the direct contribution of convection should dominate over $\mathbf{f}_\text{ST}$ in the spheroidal equation of motion when magnetic fields are weak (the coupling strength scales like the magnetic pressure $B^2$, which is far smaller than the gas pressure). We shall thus neglect all magnetic terms in Eq.~(\ref{eq:EoM_sph_lin}). In contrast, the Lorentz force has first-order significance in providing both the driving and the restoration of torsional motions, since there are no pressure or buoyancy forces to compete with, and cannot be neglected in Eq.~(\ref{eq:EoM_tor_lin_2}). 

With magnetic terms neglected, finding the spheroidal eigenmodes reduces to the standard hydrodynamic problem of linear adiabatic stellar oscillations. In the Cowling approximation (i.e.~neglecting the perturbation to the gravitational potential), this involves solving the following second-order system of ODEs:
\begin{align}
  \frac{\rmd \xi_r}{\rmd r} &= -\left( \frac{2}{r} - \frac{1}{\Gamma_1 H_p} \right) \xi_r + \frac{1}{\rho_0 c_s^2} \left( \frac{S_\ell^2}{\omega^2} - 1 \right) p' \nonumber \\
  \frac{\rmd p'}{\rmd r} &= \rho_0 (\omega^2 - N^2) \xi_r - \frac{1}{\Gamma_1 H_p} p' \:, \label{eq:osc_Cowling}
\end{align}
which can be achieved by standard numerical techniques for ODE eigenvalue problems. We did this by first interpolating the CESAM profiles onto a finer grid (to capture the small spatial scales of g-mode oscillations) and then solving Eqs (\ref{eq:osc_Cowling}) using the shooting method. The quantities $\Gamma_1$, $H_p$, $c_s$, $S_\ell$ and $N$ characterise the stellar background and correspond to the adiabatic index, pressure scale height, sound speed, Lamb frequency and buoyancy frequency, respectively. The horizontal component $\xi_h$ of the fluid displacement is related to $p'$ through
\begin{align}
  \xi_h = \frac{p'}{r \omega^2 \rho_0} \:. \label{eq:xi_h}
\end{align}

\begin{figure}
  \centering
  \includegraphics[width=\columnwidth]{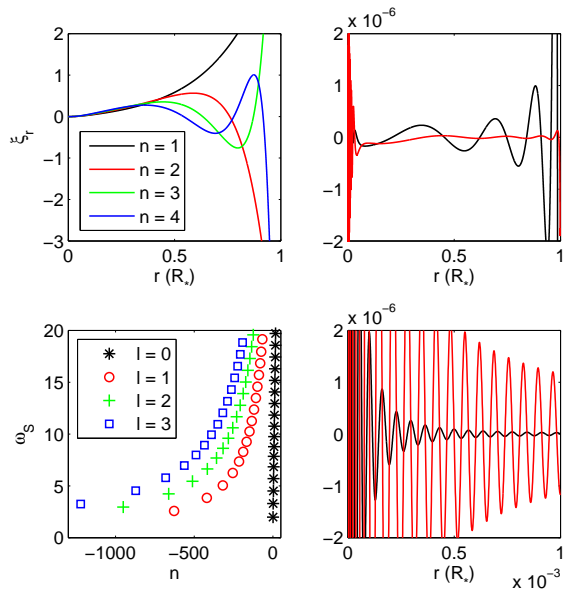}
  \caption{A selection of eigenmodes and eigenfrequencies for the stellar model examined here, showing the four lowest-order radial modes (top left), and an $\ell = 1$ mixed mode near $\omega = 10$ (top right) with the central regions shown enlarged on the bottom right so that the g-type oscillations can be seen. Red and black in the two righthand plots correspond to horizontal and radial fluid displacements $\xi_h$ and $\xi_r$, respectively. Note that the scaling of $\boldsymbol{\xi}$ is arbitrary. The frequencies (expressed as a multiple of the dynamical frequency) of the first four lowest spherical degrees are plotted versus radial order on the bottom left.}
  \label{fig:Seig}
\end{figure}

We conducted a near-exhaustive search for all spheroidal eigenmodes between $\omega = 1$ and 20 (expressed as a multiple of the dynamical frequency $\sqrt{GM_*/R_*^3}$), refining this near the locations of p-dominated modes (of which one exists per radial order), for spherical harmonic degrees $\ell = 0$, 1, 2 and 3. The objective was to selectively extract the modes that are observable experimentally, which are restricted to those with low $\ell$ (due to geometric cancellation effects), strong p-mode character (associated with larger surface motions and therefore intensity variations) and located within several radial orders of $\omega = 10$ (typical frequency of maximum excitation for solar-like oscillators). Figure \ref{fig:Seig} shows several $\ell = 0$ modes and one $\ell = 1$ mode found by the search. The $\ell = 0$ modes (top left) are oscillatory only near the surface, while the $\ell = 1$ mode (top right and bottom right) has significant mixed character and is oscillatory both near the surface and near the centre. Mixing occurs also for $\ell = 2$ and 3, but due to the weaker coupling for higher $\ell$, the modes have purer p- or g-like character compared to $\ell = 1$. Radial orders $n$ (bottom left) were computed using the Eckart scheme \citep{Eckart1960, Scuflaire1974, Osaki1975}, where the convention is to count p-type (g-type) radial crossings positively (negatively). The large negative values for the $\ell > 0$ modes indicate that these have a large number (hundreds) of oscillations in the g-mode cavity.

\begin{figure}
  \centering
  \includegraphics[width=0.9\columnwidth]{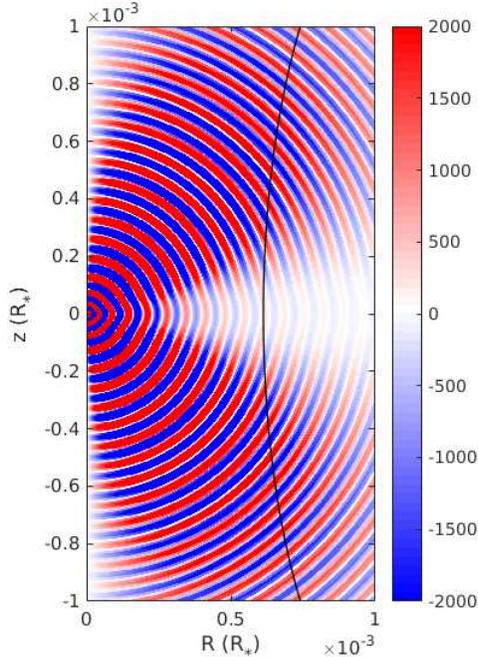}
  \caption{Spatial distribution of $f_\text{TS}$ (the torsional component of the Lorentz force associated with spheroidal motions) for the mixed mode shown on the right of Fig.~\ref{fig:Seig}, overlaid with the flux surface shown in Fig.~\ref{fig:vAmag_X150}. Only the region near the centre is shown. Fluid displacements have been normalised so that the total energy of the mode is unity. One sees that there is a cross-cut of the field loop across many radial nodes (sign changes) of $f_\text{TS}$, suggesting that this should preferentially excite high-index harmonics on that loop.}
  \label{fig:JxB}
\end{figure}

Recall that from the point of view of the torsional equation of motion, spheroidal fluid motions act as a forcing function through their associated Lorentz force. In general, if the forcing applied to a mechanical oscillator contains one or more frequencies that match its natural frequencies, then resonant excitation occurs. The strength of the excitation depends on the geometric similarity between the forcing function at the resonant frequencies and the corresponding eigenmodes, which can be quantified as the coefficients of the eigenfunction expansion of the forcing function. Figure \ref{fig:JxB} shows the spatial distribution of $f_\text{TS}$ near the core for the mode shown on the right of Fig.~\ref{fig:Seig}. Overlaid in black is the same field loop shown in Fig.~\ref{fig:vAmag_X150}. The fine-scale oscillations of $\boldsymbol{\xi}_S$ produce corresponding fine-scale oscillations of $f_\text{TS}(\boldsymbol{\xi}_S)$. This illustrates how the excitation of high harmonics by low-degree modes can occur: in general the field lines cut across many radial shells, enabling large $n$ to map to large $j$.

The torsional equation of motion in terms of the scaled fluid displacement $\eta_\phi$ can be written
\begin{align}
  \frac{\del^2 \eta_\phi}{\del t^2} - \frac{1}{\rho_0 R^4} \frac{\del^2 \eta_\phi}{\del \sigma^2} = F_\phi \:, \label{eq:EoM_tor_eta}
\end{align}
where $F_\phi = f_\text{TS} / \rho_0 R$. We now wish to derive an expression for the expansion coefficients of $F_\phi$ (the spheroidal forcing) with respect to the torsional eigenmodes $X_j$ identified in Section \ref{sec:torsional}, i.e.~the quantities $a_j$ in
\begin{align}
  F_\phi(\sigma, \psi) = \sum_j a_j(\psi) X_j(\sigma, \psi) \:. \label{eq:Fphi_expand}
\end{align} 

First, we need to establish the orthogonality relation for $X_j$. Substituting the eigensolution $\omega_{0,j}^2, X_j$ into the homogeneous form of Eq.~(\ref{eq:EoM_tor_eta}), we have
\begin{align}
  -\rho_0 R^4 \omega_{0,j}^2 X_j = \frac{\del^2 X_j}{\del \sigma^2} \:. \label{eq:EoM_tor_eta_homo}
\end{align}
Integrating twice by parts and applying periodic boundary conditions, one can show that
\begin{align}
  \oint X_k^* \frac{\del^2 X_j}{\del \sigma^2} \rmd \sigma = \oint X_j \frac{\del^2 X_k^*}{\del \sigma^2} \rmd \sigma \:, \label{eq:Teig_iden}
\end{align}
where the integral is around a closed field loop. Multiplying Eq.~(\ref{eq:EoM_tor_eta_homo}) by $X_k^*$ and integrating, and using Eq.~(\ref{eq:Teig_iden}), we find that
\begin{align}
  \left( \omega_{0,j}^2 - \omega_{0,k}^2 \right) \oint \rho_0 R^4 X_j X_k^* \rmd \sigma = 0 \:, \label{eq:Teig_ortho}
\end{align}
which implies that unless $\omega_{0,j}^2 = \omega_{0,k}^2$, it must be that $\oint \rho_0 R^4 X_j X_k^* \rmd \sigma = 0$. It is possible to normalise the $X_j$ so that $\oint \rho_0 R^4 X_j X_j^* \rmd \sigma = 1$. Doing so, we arrive at the desired expression
\begin{align}
  a_j = \oint R^3 f_\text{TS} X_j^* \rmd \sigma \:. \label{eq:a_j}
\end{align}

\begin{figure}
  \centering
  \includegraphics[width=\columnwidth]{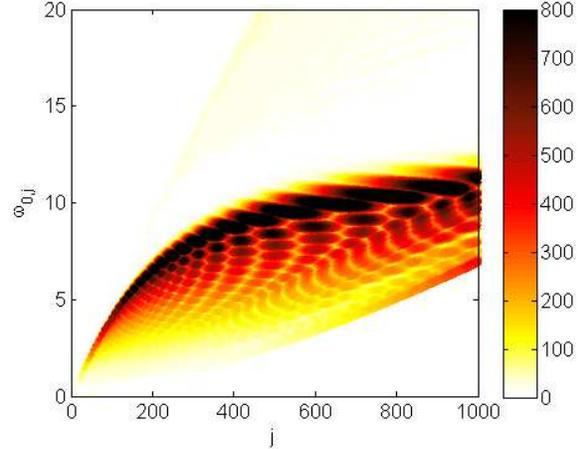}
  \caption{The torsional spectrum computed for 500 flux surfaces. Points are coloured according to the strength of coupling with the spheroidal mode whose Lorentz force distribution is shown in Fig.~\ref{fig:JxB}. The coupling strength is quantified as $|a_j|$, the absolute value of the coefficient of the eigenfunction expansion (see Eq.~(\ref{eq:a_j})).}
  \label{fig:coupling_strengths}
\end{figure}

The values of $a_j$ for the spheroidal mode shown in Fig.~\ref{fig:JxB} and each of the torsional modes are plotted in Fig.~\ref{fig:coupling_strengths}. The spheroidal displacements have been normalised such that the total energy $E$ \citep{Unno1989} equals unity, i.e.
\begin{align}
  E = \omega^2 \int \rho_0 r^2 \left[ \xi_r^2(r) + \ell (\ell + 1) \xi_h^2(r) \right] \rmd r = 1 \:. \label{eq:sph_norm}
\end{align}
As a comment, the values obtained for $a_j$ for the current model are substantially lower than the maximum physically allowed values, which would occur under conditions of complete geometric overlap (perfect constructive interference). As an order-of-magnitude estimate, the upper bound on $a_j$ is given by $R_c^{1/2} B^{3/2} \xi L^{-2}$, where $R_c$ is the size of the core, $\xi$ is the characteristic fluid displacement and $L$ is the characteristic length scale of variation in $\xi$. For our red giant model, substituting appropriate values for these parameters yields a physical upper bound on $a_j$ of the order $10^4$. As can be seen from Fig.~\ref{fig:coupling_strengths}, the actual values obtained are $a_j \sim 100$.

Also apparent from Fig.~\ref{fig:coupling_strengths} is that the torsional spectrum is very dense. In fact, for every value of $j \in \mathbb{Z}^+$ there is a continuum of $\omega_{0,j}$ values, reflecting the existence of a continuum of flux surfaces. However, the number of resonances is finite due to the discrete nature of $j$, and at a given $\omega$ equals the $j$ range intersected by a horizontal line at that value of $\omega$. For the spheroidal mode shown in Fig.~\ref{fig:JxB}, which is near $\omega = 10$, we identify around 900 resonances, i.e.~there exist this number of magnetic flux surfaces which have a torsional mode with this eigenfrequency.

\subsection{Dissipative effects}\label{sec:dissipation}
We shall now incorporate dissipative effects, with the goal of eluciating their role in contributing to the damping of the torsional oscillations. In the following derivation it will be assumed that the damping coefficients in units of the dynamical frequency are much less than unity (highly underdamped oscillator), so that the eigenfrequencies and eigenmodes derived above remain valid. At the end of this section we evaluate the expression for the damping coefficient obtained and verify that it is indeed small.

With dissipative terms included, the momentum and induction equations are
\begin{align}
  \frac{\del^2 \boldsymbol{\xi}}{\del t^2} + \mathcal{L}[\boldsymbol{\xi}] - \nu \nabla^2 \frac{\del \boldsymbol{\xi}}{\del t} &= 0 \label{eq:EoM_diss} \\
  \frac{\del \mathbf{B}'}{\del t} - \nu_m \nabla^2 \mathbf{B}' &= \nabla \times \left( \frac{\del \boldsymbol{\xi}}{\del t} \times \mathbf{B}_0 \right) \:, \label{eq:ind_diss}
\end{align}
where $\nu$ and $\nu_m$ are the viscous and Ohmic dissipation coefficients, and $\mathcal{L}[\boldsymbol{\xi}]$ refers to the linearised form of the RHS of Eq.~(\ref{eq:EoM}). Anticipating small scales we have retained only the highest order derivatives in the viscous force and rate of Ohmic diffusion. Invoking a time-harmonic separation, Eq.~(\ref{eq:ind_diss}) becomes an inhomogeneous Helmholtz equation which can be solved by means of an integration kernel to yield $\mathbf{B}'(\mathbf{r}) \approx \nabla \times (\bar{\boldsymbol{\xi}} \times \mathbf{B}_0)$, where
\begin{align}
  \bar{\boldsymbol{\xi}}(\mathbf{r}) &= \iiint K(\mathbf{r} - \mathbf{r}') \boldsymbol{\xi}(\mathbf{r}') \rmd^3 \mathbf{r}' \:, \label{eq:deloc_xi} \\
  K(\mathbf{r}) &= \frac{\rmi \omega}{4\pi \nu_m |\mathbf{r}|} \exp \left[ -(1-i) \sqrt{\frac{\omega}{2\nu_m}} |\mathbf{r}| \right] \:. \label{eq:diss_kernel}
\end{align}
Here we have assumed that $\boldsymbol{\xi}$ varies much more rapidly over space than $\mathbf{B}_0$, which itself varies on a scale much larger than $\sqrt{\nu_m/\omega}$. As far as the Lorentz force is concerned the effect of dissipation is thus to replace $\boldsymbol{\xi}$ by $\bar{\boldsymbol{\xi}}$, which are related through $\boldsymbol{\xi} = \bar{\boldsymbol{\xi}} - (\nu_m/\rmi \omega) \nabla^2 \bar{\boldsymbol{\xi}}$. Substituting this into Eq.~(\ref{eq:EoM_diss}), neglecting products of $\nu$ and $\nu_m$ (given that both are small), defining $\bar{\eta}_\phi \equiv \bar{\xi}_\phi/R$ and retaining only diffusive terms involving second order spatial derivatives, the torsional component can be written
\begin{align}
  \frac{\del^2 \bar{\eta}_\phi}{\del t^2} - \frac{1}{\rho_0 R^4} \frac{\del^2 \bar{\eta}_\phi}{\del \sigma^2} - \nu_\text{tot} \frac{\del^2}{\del n^2} \frac{\del \bar{\eta}_\phi}{\del t} = F_\phi \:, \label{eq:EoM_tor_diss}
\end{align}
where $\nu_\text{tot} = \nu + \nu_m$ is the total dissipation coefficient and $n$ is the direction normal to the flux surfaces. The reason for retaining only this spatial part of the Laplacian is that the finest-scale structure is likely to develop in the direction perpendicular rather than parallel to the flux surfaces, as a result of phase mixing. This refers to the decorrelation of oscillations on adjacent surfaces having slightly different eigenfrequencies, which occurs on a timescale corresponding to the inverse of their frequency difference (cf.~the beat phenomenon). In the case of our model, the decorrelation timescale associated with the spatial scale along flux surfaces ($\sim 10^{-5} R_*$) is only $\sim$10 dynamical times, much shorter than the dissipation timescale associated with the same length scale ($\sim 10^6$ dynamical times). Decorrelation will therefore proceed down to much smaller scales before its development is halted by viscous/resistive effects.

We seek a solution to the coefficients $b_j$ of the eigenfunction expansion $\bar{\eta}_\phi(t,\sigma,\psi) = \sum_j b_j(\psi) X_j(\sigma,\psi) \rme^{-\rmi \omega t}$. Noting that $\del/\del n = RB_p \del/\del \psi$, we get
\begin{align}
  \left[ \omega_{0,j}^2 - \omega^2 \right] b_j(\psi) + \rmi \nu_\text{tot} \omega R^2 B_p^2 \frac{\del^2 b_j(\psi)}{\del \psi^2} = a_j(\psi) \:. \label{eq:EoM_tor_diss_coeff}
\end{align}
Let us focus on a small region in $\psi$ near a resonant surface $\psi_0$ whose $j$-th harmonic is of frequency $\omega_{0,j}(\psi_0) = \omega$. Locally, we adopt the Taylor expansion $\omega_{0,j}^2(\psi) \approx \omega^2 + 2\omega\omega_{0,j}'(\psi_0) [\psi - \psi_0]$. Consider the change of variable $x = C[\psi-\psi_0]$ where $C = (2\omega_{0,j}'(\psi_0)/R^2 B_p^2)^{1/3}$, which turns Eq.~(\ref{eq:EoM_tor_diss_coeff}) into
\begin{align}
  b_j x + \rmi \nu_\text{tot} \frac{\del^2 b_j}{\del x^2} = \frac{C a_j(\psi_0)}{2\omega\omega_{0,j}'(\psi_0)} \:. \label{eq:EoM_tor_diss_coeff_loc}
\end{align}
The RHS, which we will call $A$, can be regarded as roughly constant over the resonant layer. Equation (\ref{eq:EoM_tor_diss_coeff_loc}) can be solved using Fourier transforms ($x \to k$ and $b_j(x) \to \tilde{b}_j(k)$), yielding
\begin{align}
  \tilde{b}_j(k) = \begin{cases}
    0 & k > 0 \\
    \rmi A \exp[k^3 \nu_\text{tot} / 3] & k < 0
  \end{cases} \label{eq:b_jk}
\end{align}
In the limit $\nu_\text{tot} \to 0$, $\tilde{b}_j/\rmi A$ tends to $1-H(k)$ where $H(k)$ is the Heaviside step function. This satisfies $\rmi \mathcal{F}[1-H(k)] = 1/(x - \rmi 0)$. Here $\mathcal{F}$ denotes a Fourier transform and $\rmi 0$ is an infinitesimal imaginary component that we identify with the damping contribution $\rmi \Gamma \omega$ (cf.~the Landau prescription from plasma physics). The final solution for $b_j$ is then
\begin{align}
  b_j(\psi) = \frac{a_j(\psi_0)}{\omega_{0,j}^2(\psi) - \omega^2 - \rmi \Gamma \omega} \:. \label{eq:b_j}
\end{align}
Since the RHS of Eq.~(\ref{eq:EoM_tor_diss_coeff_loc}) is approximately constant, we infer the characteristic scale to be $x \sim \nu_\text{tot}^{1/3}$. The local damping rate can be estimated from the first-order term of the Taylor expansion of $\omega_{0,j}^2(\psi)$, and has the approximate expression $\Gamma \sim [2\omega_{0,j}'(\psi_0) RB_p]^{2/3} \nu_\text{tot}^{1/3}$. Up to a factor of order unity, $\Gamma$ turns out to be the inverse of the timescale required for decorrelation to occur over the width of the resonant layer. Further inspection reveals that this width is precisely that for which the timescales of decorrelation and dissipation are equal, providing the physical interpretation for the associated loss process as being closely linked to phase mixing. Given the small magnetic Prandtl numbers in stellar interiors, $\nu_\text{tot}$ is dominated by the Ohmic dissipation coefficient $\nu_m \approx 10^9 T^{-3/2}$\,m$^2$\,s$^{-1}$ \citep{Spitzer1962}. For our model ($T \sim 10^7$\,K) we find that $\Gamma \sim 10^{-3}$ inverse dynamical times, and that the widths of the resonant layers are $\sim 10^{-7} R_*$.

\subsection{Overall damping rates}\label{sec:overall_damping}
A major objective of this work is to estimate the overall damping rate $\gamma$ of spheroidal modes due to the resonant coupling with torsional modes. We will now combine results from preceding sections to arrive at an expression for $\gamma$.

The total rate of work done by the torsional component of the Lorentz force associated with spheroidal motions is
\begin{align}
  \frac{\rmd E}{\rmd t} &= \iiint \left( f_\text{TS} \frac{\del \xi_\phi^*}{\del t} + f_\text{TS}^* \frac{\del \xi_\phi}{\del t} \right) \rmd^3 \mathbf{r} \nonumber \\
  &= 4\pi \text{Re}\left[ \iint \rho_0 R^4 F_\phi \frac{\del \eta_\phi^*}{\del t} \rmd \sigma \rmd \psi \right] \:. \label{eq:dEdt}
\end{align}
Invoking the eigenfunction expansions for $F_\phi$ and $\eta_\phi$, eliminating $b_j$ in favour of $a_j$ using Eq.~(\ref{eq:b_j}), making use of the orthonormality relation for $X_j$ and averaging over one oscillation period, we obtain the time-averaged rate of work
\begin{align}
  \bigg\langle \frac{\rmd E}{\rmd t} \bigg\rangle &\approx 4\pi \sum_j \left| a_j(\psi_0) \right|^2 \int \frac{\Gamma}{h(\psi) + \Gamma^2} \rmd \psi \:, \label{eq:dEdt_tav}
  \shortintertext{where}
  h(\psi) &\equiv \left( \frac{\omega_{0,j}^2(\psi)}{\omega} - \omega \right)^2 \:. \label{eq:h}
\end{align}
This assumes that $a_j$ varies slowly over the width of the resonant region and can be approximated by its value at the resonant surface $\psi = \psi_0$ where $\omega_{0,j}(\psi_0) = \omega$. In the limit of small $\Gamma$ the resonant region is spatially narrow, and so we can approximate $h(\psi)$ by the first term of its Taylor expansion about $\psi_0$. This allows us to straightforwardly evaluate the $\psi$-integral in Eq.~(\ref{eq:dEdt_tav}). We arrive at the final expression
\begin{align}
  \bigg\langle \frac{\rmd E}{\rmd t} \bigg\rangle = 2\pi^2 \sum_j \left| a_j(\psi_0) \right|^2 \left( \left| \frac{\rmd \omega_{0,j}}{\rmd \psi} \right|_{\psi_0} \right)^{-1} \:, \label{eq:dEdt_final}
\end{align}
which we see is independent of the local damping coefficient $\Gamma$. This reflects a basic property of driven-damped oscillators that near a resonance, in the limit of weak dissipation (regardless of what this may be or how it physically arises), the system always adjusts itself so that the rate of driving and dissipation are in balance. The global damping rate of a spheroidal mode is then $\gamma = \langle \rmd E/\rmd t \rangle / E$, where $E$, given in Eq.~(\ref{eq:sph_norm}), is the total energy of the mode. If we were to normalise the fluid displacements such that $E = 1$, then $\gamma$ is simply given by Eq.~(\ref{eq:dEdt_final}).

\section{Results}\label{sec:results}
\subsection{Application: damping red giant oscillations}
The summation in Eq.~(\ref{eq:dEdt_final}) for any given spheroidal mode having frequency $\omega$ is over all torsional modes (indexed by $j$) whose frequencies equal this. Our approach to evaluating this was to scan over the $1000 \times 1000$ torsional modes computed and saved for the ones whose $\omega_{0,j}$ lay closest to the target frequency. If the difference fell below a certain threshold (taken to be 0.1 frequency units), this was classed as a resonance and the contribution of the mode was added to the sum. The typical $\omega$ spacing between the torsional modes in our set was $\sim$0.01 frequency units, so variations about the threshold of 0.1 units affected the results very little. The purpose of the threshold was to restrict the $j$-range of modes resonant at a given $\omega$ to be as close to the true range as possible, without losing modes comfortably within the true range. A further point to be made is that every $j$ value was scanned independently, which restricts the resulting set of modes to one per $j$ but not one per flux surface. At the low resolution of the grid (1000 flux surfaces) compared to the $j$-range of the spectrum (of order this value, as can be seen from Fig.~\ref{fig:coupling_strengths}), there are instances where a flux surface contributes more than one mode to the sum. Though not strictly realistic, it is inevitable with this discretised approach that the true resonant surface for given $j$ is approximated by one close by, and this may be shared between more than one $j$. Given the slow variation in the structure of the torsional modes with space, however, this is unlikely to give rise to systematic errors. Our quantitative results are impacted more heavily by the fact that only a finite number of modes were generated, since this excludes some number of potentially resonant modes (implications are discussed further in Section \ref{sec:limits}).

\begin{figure}
  \centering
  \includegraphics[width=\columnwidth]{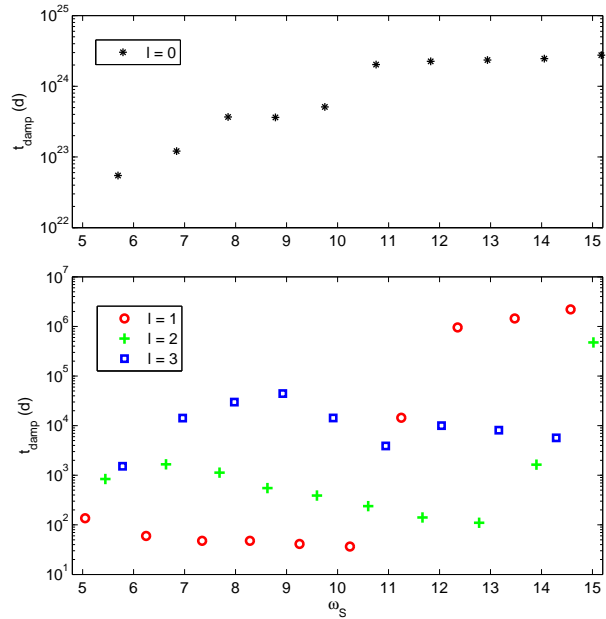}
  \caption{Damping times for the most p-dominated modes of the four lowest-degree spherical harmonics. Radial modes are shown on a separate plot (top) since their damping times differ greatly from the non-radial modes (bottom). For non-radial modes, at frequencies below a certain threshold (here $\omega \approx 11$), one sees that lower degrees experience stronger damping. Note the logarithmic scale on the vertical axis. The unit of $\omega_S$ is the dynamical frequency, while $t_\text{damp}$ is expressed in days.}
  \label{fig:tdamp_vs_om}
\end{figure}

Figure \ref{fig:tdamp_vs_om} plots the damping times $t_\text{damp} \equiv \gamma^{-1}$ calculated for the p-dominated modes of the red giant model described in Section \ref{sec:models}. The field strength is scaled such that the central value is 4\,MG, in line with expectations from simulations of main sequence core dynamos \citep{Brun2005} and magnetic flux conservation during core contraction into the red giant stage. We find that damping times for the $\ell = 0$ modes are typically in excess of $10^{23}$ days, many orders of magnitude longer than for the $\ell > 0$ modes, which are damped on timescales of months through this mechanism. The distinguishing feature between radial and non-radial modes which is likely to explain this is the structure of the spatial amplitude function within the core. For radial modes, spatial oscillations occur only near the surface; near the core the amplitude function follows a very smooth exponential decay. This results in very tiny $a_j$ values for the torsional modes (which are localised to the core) having frequencies near $\omega = 10$. In contrast, the spatial amplitude functions of non-radial modes near $\omega  = 10$ have hundreds of finely-spaced oscillations within the core, and so are able to strongly overlap with torsional modes having $j$ values of several hundred. At the field strengths expected for red giant cores, it turns out that the eigenfrequencies associated with torsional modes having $j \sim 100$ lie in the vicinity of $\omega = 10$, making resonances between spheroidal and torsional modes possible. (For $\rho \sim 10^5$\,g\,cm$^{-3}$, $B \sim 1$\,MG and $R_c \sim 10$\,Mm we have $v_A \sim 10$\,m\,s$^{-1}$, implying a fundamental Alfv\'{e}n frequency of $\sim$1\,$\umu$Hz. Spheroidal modes in red giants are excited near $\sim$100\,$\umu$Hz.)

\subsection{Dependence on spherical harmonic degree}
For the three non-radial degrees examined we find systematic differences in the characteristic damping times, these differing by roughly an order of magnitude between adjacent $\ell$. At the field strength considered, damping times below a certain threshold frequency (occurring e.g.~near $\omega = 11$ for $\ell = 1$; this ``step'' feature is commented on more below) are $10^1$--$10^2$\,days for $\ell = 1$, $10^2$--$10^3$ days for $\ell = 2$, and $10^3$--$10^5$ days for $\ell = 3$. Particularly for $\ell = 1$ these damping times are comparable to those associated with turbulent convection (10--30 days), which suggests that damping through resonant interactions with Alfv\'{e}n modes should impact mode amplitudes at an observable level. These would be most pronounced for $\ell = 1$, followed by $\ell = 2$ and then $\ell = 3$. 

\begin{figure}
  \centering
  \includegraphics[width=\columnwidth]{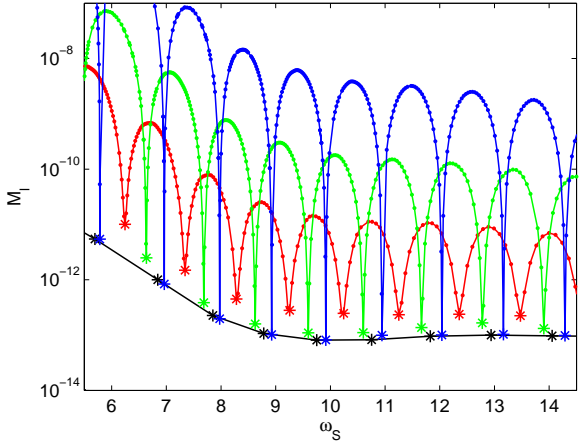}
  \caption{Mode inertia as a function of frequency for the spheroidal modes of the red giant model. Black, red, green and blue correspond to $\ell = 0$, 1, 2 and 3, respectively. Though the spectrum is discrete, individual points (one for each mode) have been joined by straight lines to aid visualisation of the pattern. Asterisks mark the most p-dominated modes, identified as local minima in the red, green and blue curves. These are the ones for which damping times have been computed and plotted in Fig.~\ref{fig:tdamp_vs_om}.}
  \label{fig:Ml_vs_om}
\end{figure}

The physical reason for the dependence of damping rates on $\ell$ is likely to be the variation in the strength of coupling between the p- and g-mode cavities, which determines the extent of mode mixing. For a fixed frequency the g-mode cavity is of the same size regardless of $\ell$, but the p-mode cavity is larger for smaller $\ell$. Consequently mode mixing is most effective for $\ell = 1$, followed by $\ell = 2$, then $\ell = 3$, and so on for higher multipoles. One measure of the g-like character of a mode is the mode inertia
\begin{align}
  M_\ell = \frac{\int \rho_0 r^2 \left[ \xi_r^2(r) + \ell(\ell+1) \xi_h^2(r) \right] \rmd r}{\xi_r^2(R_*) + \ell(\ell+1) \xi_h^2(R_*)} \:, \label{eq:Ml}
\end{align}
which gives an indication of the amount of mass displaced by the mode. Modes with smaller (larger) inertia are more p-like (g-like), preferentially localised to the envelope (core) where densities are lower (higher). The plot of $M_\ell$ versus mode frequency (see Fig.~\ref{fig:Ml_vs_om}) exhibits periodic modulations once per radial order (one unit of $\omega$), where the modes having minimum $M_\ell$ are the most p-dominated ones. Observations favour the detection of p-dominated modes because these give rise to larger fluid motions at the surface, and so we selectively consider these modes here. These are marked with asterisks in Fig.~\ref{fig:Ml_vs_om}. One can see that the $\ell = 1$ p-dominated modes (red asterisks) have the largest inertia and thus the most g-like character compared to other $\ell$. The larger core fluid displacements associated with the p-dominated $\ell = 1$ modes enhances their rate of damping due to interactions with the torsional Alfv\'{e}n modes, compared with the p-dominated modes of higher $\ell$ (green and blue asterisks).

\subsection{Scaling with field strength and core size}\label{sec:scaling}
For a fixed core size $R_c$, damping rates associated with this mechanism are expected to be smaller for weaker magnetic fields. We have $f_\text{TS} \propto B^2$, $X_j \propto B^{1/2}$ and $\oint \rmd \sigma \propto B^{-1}$, so if we were to ignore changes in $a_j$ associated with details of the mode geometries, then from Eq.~(\ref{eq:a_j}) we infer that $a_j \propto B^{3/2}$. In addition the number of available resonances at given $\omega$ is inversely proportional to $B$, and $|\rmd \omega_{0,j}/\rmd \psi|$ is independent of $B$, so from Eq.~(\ref{eq:dEdt_final}) this implies that $\gamma \propto B^2$. The increase in $\gamma$ with $B$, which appears to be driven through the $a_j$ dependence, arises physically from the increased coupling between spheroidal and torsional motions when the Lorentz force is stronger. While the above scaling argument is fairly simplistic, it does appear to be the case numerically that $\gamma$ increases with increasing $B$ (doubling the field strength decreases the characteristic damping times by a factor of several).

Also of interest is the predicted variation of $\gamma$ with $R_c$. As a star ascends the red giant branch its core contracts and envelope expands. One expects the damping rate of modes near the frequency of maximum excitation $\nu_\text{max}$ (which is itself subject to variation) to change accordingly. To estimate this we will assume conservation of mass and magnetic flux, so that $\rho_0 \propto R_c^{-3}$ and $B \propto R_c^{-2}$. In this situation $f_\text{TS} \propto R_c^{-5}$, $\oint \rmd \sigma \propto R_c$ and $X_j \propto R_c^{-1}$, so $a_j \propto R_c^{-2}$. We also need to account for changes in the torsional eigenfrequencies: these go as $\omega_{0,j} \sim v_A/R_c$ where $v_A \sim B/\sqrt{\rho_0} \propto R_c^{-1/2}$, so both $\omega_{0,j}$ and $|\rmd \omega_{0,j}/\rmd \psi| \propto R_c^{-3/2}$. 

To predict the dependence of $\gamma$ on $R_c$, it remains to determine the effect of changing the frequency of excitation, which enters into $\gamma$ through $\sum_j$, the summation over resonant modes. As can be seen from Fig.~\ref{fig:coupling_strengths}, the torsional eigenspectrum fills a wedge in $(\omega_{0,j},j)$-space. The number of resonances at frequency $\omega$ equals the range in $j$ intersected by a horizontal line placed at that frequency. If all $\omega_{0,j}$ are boosted by a given factor, then the wedge is stretched upward leaving a proportionally smaller $j$-intersection range at the location of a fixed horizontal line. Likewise, if the line representing $\omega$ is shifted downwards by some factor, then the $j$-range intersected decreases proportionally. Inspection of the CESAM models for this star shows that between 963 and 1021\,Myr, $R_*$ roughly quadruples while $R_c$ halves, suggesting that empirically $R_* \propto R_c^{-2}$. We know that $\nu_\text{max} \propto R_*^{-3/2}$, so putting all this together, $\sum_j \propto \nu_\text{max} / \omega_{0,j} \propto R_c^{9/2}$. Together with the $|a_j|^2 |\rmd \omega_{0,j}/\rmd \psi|^{-1} \propto R_c^{-5/2}$ dependence of the contribution from each mode, this leads to $\gamma \propto R_c^2$ as the predicted scaling dependence for spheroidal modes near $\nu_\text{max}$. We thus predict that damping rates should \textit{decrease} as the core contracts. The dominating influence here is the reduction in the number of available resonances, which falls off more quickly than the strength of spheroidal-torsional coupling grows (acting alone, the latter would tend to drive up $\gamma$ through the increase in $B$ and therefore $a_j$ as $R_c$ shrinks).

\section{Discussion}\label{sec:discuss}
\subsection{Comparison with observations}
\subsubsection{Impact on mode visibilities}
Mode visibilities (i.e.~amplitudes) $v_\ell$, which are a measure of the area under the peaks in the power spectra associated with a given $\ell$, are usually expressed in a form where they are normalised with respect to some other quantity. If this is with respect to the area under $\ell = 0$ peaks, then for example $v_1 \approx 1.54$ \citep{Ballot2011}. In the context of the dipole dichotomy problem, following \citet{Mosser2016}, it is more convenient to normalise with respect to the area under the $\ell = 1$ peaks of stars which fall in the high-amplitude (``normal'') group. Normal stars thus have $v_1 \approx 1$. For a fixed rate of excitation, the visibility of a mode decreases proportionally with the overall damping rate, and so the latter definition of $v_1$ allows it to be expressed in terms of the envelope- and core-associated damping rates $\gamma_e$ and $\gamma$ as 
\begin{align}
  v_1 = \frac{\gamma_e}{\gamma_e + \gamma} = \frac{t_\text{damp}}{t_e + t_\text{damp}} \:, \label{eq:v1}
\end{align}
where $t_e \equiv \gamma_e^{-1}$. In practice, $t_e$ can be measured from the linewidths of the radial modes and is characteristically $15 \pm 5$\,days \citep[see][fig.~3b]{Mosser2016}

As can be seen from fig.~3a of \citet{Mosser2016}, the low-amplitude group of red giants have $v_1$ values close to 0.1 for stars with dynamical frequencies $\Delta \nu$ near 12\,$\umu$Hz, and $v_1$ values around 0.7 for stars with $\Delta \nu$ near 4\,$\umu$Hz, with some scatter about this trend. Let us take $t_e$ to be 15\,days. From Fig.~\ref{fig:tdamp_vs_om} it appears that $t_\text{damp}$ lies between 30 and 50 days for $\ell = 1$, producing $v_1$ of 0.67--0.77. Increasing the field strength increases the damping and lowers the visibilities: at triple the field strength, $t_\text{damp}$ values would be roughly nine-fold lower producing $v_1$ of 0.18--0.27. The full range of observed visibilities for stars with depressed $\ell = 1$ modes can thus be accounted for through modest variations of the field strength. For $\ell = 2$, $t_\text{damp}$ values are about an order of magnitude greater than $\ell = 1$, placing $v_2$ (defined in an analogous manner) at around 0.93--0.97. At triple the field strength, $v_2$ would be in the range 0.60--0.88. The damping times computed for $\ell = 3$ are about an order of magnitude larger still, which would give $v_3$ values of 0.98--0.99 (this would be much more difficult, if not impossible, to detect compared to $\ell = 1$ and 2). Hence the $\ell$-dependence of the mode amplitude depression observed experimentally \citep{Stello2016a} is reproduced by our mechanism.

\subsubsection{Explaining the dichotomy}
Given that $\gamma$ depends on $B$, one possible explanation for the dichotomy in the red giant population is that this reflects a dichotomy in the field strengths. This could be related to the stability of magnetic equilibria, where the initial dynamo-generated field relaxes into one of two or more possible states characterised by different equilibrium strengths \citep[e.g.][]{Braithwaite2008}. The helicity (aproximately conserved during relaxation), which in turn depends on the dynamo mechanism, may play a role. However, further investigation along this line is beyond the scope of this work.

Another possibility is that this relates to a property of the results not yet discussed in much detail, and that is the dramatic step-like feature present in the $a_j$ distribution. This refers to the broad ``shoulder'' tracing out a roughly horizontal line past about $j \gtrsim 400$ near $\omega = 11$ in Fig.~\ref{fig:coupling_strengths}. Just below this line $a_j$ values are large, but step down by about two orders of magnitude above it. This does not appear to be associated with any particular flux surface; rather, it seems as though each flux surface has a $j$ value above which $a_j$ suddenly becomes small, and the associated $\omega_{0,j}$ is roughly the same for all flux surfaces. We notice that the position of this step migrates downward for spheroidal modes of increasing frequency $\omega_S$, and so we suggest that it may be related to the spatial scale $L$ of the g-mode oscillations, which given a characteristic $v_A$ are associated with a certain frequency $v_A/L$. As $\omega_S$ increases, so does $L$, which would qualitatively explain the direction of its migration. One could conceive of this as being related to the condition $\omega_S = v_A/L$ describing a match of both the frequency and spatial scale of spheroidal and torsional modes, which gives rise to optimal coupling. Torsional modes with smaller spatial scales rapidly become difficult to excite. The step feature can be clearly seen in Fig.~\ref{fig:tdamp_vs_om} as the strong jump in $t_\text{damp}$ near $\omega = 11$ for $\ell=1$, and $\omega = 14$ for $\ell = 2$ (for $\ell = 3$ this lies slightly off the edge of the plot, near $\omega = 16$). Hence there is also a dependence on $\ell$ of the step location, this being at higher frequencies for larger $\ell$.

Regardless of the origin of this feature, if it turns out to be ubiquitous among different stellar models then this could explain the dichotomy as being created by $\nu_\text{max}$ lying above or below the step. If $\omega_S$ (the frequency of a given spheroidal mode) lies below the step, $a_j$ and hence $\gamma$ values will be high. As $\omega_S$ increases, there will come a point where it meets the frequency of the step and $a_j$ values strongly drop. Modes of higher $\omega_S$ would be subject to much weaker damping. The location of the step depends also on the field strength: for stronger fields (larger $v_A$) this occurs at higher frequencies. Though there may be other factors involved, this means that an intrinsic spread of field strengths among the red giant population could produce the dichotomy. It is worth noting that this picture implies the possibility of stars for which the step lies near $\nu_\text{max}$, about which modes of several radial orders are usually detectable. For such stars, one expects to see low visibilities at frequencies below the step, and high (``normal'') visibilities above. Interestingly, such stars have indeed been reported in the literature: at least three are known \citep{Garcia2014_mn2e, Mosser2016}. 

\subsubsection{Dependence on evolutionary stage}
As mentioned earlier in this section, observed values of $v_1$ are noted to increase as the dynamical frequency $\Delta \nu$ decreases. Variations in $\Delta \nu$ are closely tied to evolutionary stage, since expansion of the envelope as the star ascends the RGB causes $\Delta \nu$ to drop. Accompanying the evolution is a contraction of the core, which we previously argued in Section \ref{sec:scaling} is associated with a drop in $\gamma$. The predicted scaling is roughly $\gamma \propto R_c^2$, so that a star with $v_1 \approx 0.45$ at $\Delta \nu = 12 \umu$Hz would end up with $v_1 \approx 0.6$ at $\Delta \nu = 4 \umu$Hz. Although somewhat shallower than the trend seen observationally, a large number of approximations have been used, some of which may be questionable. For example, if it turns out that contraction of the core allows the field configuration to relax further (so that $\psi$ is not in fact conserved), this would weaken the dependence of $a_j$ on $R_c$ and steepen the rise in $v_1$ towards smaller $\Delta \nu$. However, it is encouraging that the simple scaling dependencies predicted by our mechanism qualitatively reproduce this aspect of the observations.

\subsection{Limitations}\label{sec:limits}
The exact quantitative values presented here are clearly sensitive to the background model, and although we have only considered one stellar model and field configuration, we have endeavoured to use ones that are as realistic as possible. The current work serves mainly to illustrate the viability of our mechanism for producing damping rates that are comparable to other known sources of damping (e.g.~convection), given reasonable field strengths. More detailed investigations of parameter space and the examination of different stellar models would be the subject of future work.

Throughout we have considered only axisymmetric spheroidal and torsional modes, where the axis of symmetry matches that of the background field. This has been convenient for the purposes of the analytic treatment here. We do not expect the generalisation to non-axisymmetry to adversely impact our results. The quality of the geometric overlap between the non-radial spheroidal modes and the torsional Alfv\'{e}n modes owes to the cross-cut of field lines across a large number of fine-scale spatial oscillations, which for low-degree spheroidal modes is relatively unaffected by angular orientation. The description of torsional modes in the non-axisymmetric case is complicated by the involvement of motions in the poloidal as well as the toroidal direction. However, in the limit of small poloidal scales and small $m$ (azimuthal order), it can be shown that the non-axisymmetric torsional eigenfunctions closely resemble those found for the axisymmetric case in that they are dominated by $\phi$ rather than $\theta$-displacements. We therefore expect them to physically interact with the non-axisymmetric spheroidal modes in a similar way to the axisymmetric case presented here, although the mathematical treatment would be less trivial. A more detailed investigation of non-axisymmetric effects will be deferred to future work. 

We have assumed linearity of the fluid motions, even though the smallness of the damping coefficient $\Gamma$ suggests that large limiting amplitudes of torsional motion may be attained. If reaching nonlinear amplitudes, one might expect wave breaking to occur. Further work to investigate the complications of nonlinearity has yet to be performed.

We used a particular magnetic field configuration (the Prendergast solution) which was convenient to implement since it can be written down in closed form. We acknowledge that it is only one possible solution to the Grad-Shafranov equation; other equilibria may be permitted in reality. However the geometry of the field is not particularly important to our mechanism, because for the same characteristic Alfv\'{e}n speed, the frequencies of potentially resonant torsional modes are set by the spatial scale of the g-mode oscillations. This has no bearing on the length or the shape of the field loops. Notice that the typical conditions (field strengths, densities etc.) in red giant cores are such that the ratio of the Alfv\'{e}n speed to the g-mode wavelength fortuitously coincides with the frequency of maximum excitation $\nu_\text{max}$ of the spheroidal modes. Resonant interactions proceed effectively for this reason (an alternate way of understanding this is the requirement that gravity wave phase speeds match the Alfv\'{e}n speed). This should also be true of different field configurations, including ones for which the spatial scale of the field loops may be much smaller, as long as core field strengths are similar across the red giant population.

The only other source of damping considered here, besides that of our proposed mechanism, is that arising from convection. In reality, radiative diffusion also contributes to the damping of g-mode oscillations. We have done some rough estimates of the expected rate of damping from radiative diffusion and find that it is several orders of magnitude smaller than that associated with our proposed mechanism (for $\ell = 1$ and 2; they may be on par for $\ell = 3$). This is in line with previous works, which have determined that radiative damping is small compared to convective damping for p-dominated modes and have difficulty accounting for the low dipole mode amplitudes seen red giants \citep{Dupret2009_mn2e, Garcia2014_mn2e}; it also offers no explanation for the dichotomy.

The primary source of systematic error in our quantitative estimates of $\gamma$ arises from the finiteness of the grid. This impacts the calculation in two ways. Firstly, the torsional modes considered here were limited to those with $j \leq 1000$, which means that any resonant modes with $j > 1000$ are lost to the sum in Eq.~(\ref{eq:dEdt_final}). Importantly, this means that the damping rates presented here are likely to be systematic \textit{underestimates}, perhaps by a few tens of per cent. A second way in which $\gamma$ may be affected is by inaccuracies in the shapes of the torsional modes when $j$ is comparable to the number of grid points. Tests comparing the eigenspectra generated under different grid resolutions indicate that inaccuracies become substantial for $j$ values in excess of about half the number of grid points (discrepancies in $\omega_{0,j}$ of order unity are reached, when compared to a five-fold increase in the number of grid points). If an eigenfunction is not resolved properly its $a_j$ values will not be accurate, and it is not straightforward to predict whether systematically higher or lower values would be obtained. For the sake of caution, we restricted the largest $j$-value considered to be one-fifth of the number of grid points. The overall error in our damping rates is thus likely to be dominated by the former effect.

Finally, a key assumption is that the structure of the spheroidal modes is unaffected by the magnetic field. Clearly if they were to be modified significantly in the region of the core, then this would impact our quantitative results. We stress that our mechanism, as presented, is designed to operate in the weak-field regime where this assumption holds. The work of \citet{Fuller2015} and \citet{Lecoanet2016} has suggested that drastic alteration to spheroidal mode structure is to be expected if the field exceeds a critical strength, and so we expect that in this regime our mechanism would break down, or require modification. We therefore address a complementary regime to that of the above authors. On this note, the significance of this work is that it demonstrates a new mechanism for damping stellar oscillations that does not need to disrupt the structure of the modes in any part of the star. This is precisely what is required to account for the existence of the weak-amplitude mixed modes reported by \citet{Mosser2016}.

\section{Summary and future work}\label{sec:conclude}
We have presented a mechanism for damping spheroidal modes of red giant stars via resonant interactions with torsional Alfv\'{e}n modes localised to a magnetised core. Quantitative estimates of the associated damping rates indicate that these can be comparable to those of envelope-based sources. This may be a viable answer to the dipole dichotomy problem in the case of weak core fields, where the structure of the modes is expected to be preserved. To our knowledge, there is no other mechanism which has been proposed that achieves this for the weak-field regime.

Our mechanism can be summarised as follows. Turbulent convection in the envelope excites the usual spheroidal modes restored by pressure and buoyancy. If there is a magnetic field present inside the core, then this allows in addition for the existence of torsional oscillations restored by magnetic tension. These torsional modes can be thought of as quantised vibrations on closed field loops (standing Alfv\'{e}n waves). Coupling to spheroidal motions proceeds via the Lorentz force, since in general the Lorentz force associated with spheroidal fluid displacements has a component in the torsional direction (and vice versa). Though the fundamental frequencies of torsional modes are small for realistic field strengths, resonances with spheroidal modes are possible through the excitation of high loop harmonics. The strength of the interaction is determined by the quality of the geometric overlap between the two types of modes. In red giants, spheroidal eigenmodes have extremely fine-scale radial structure in the core, providing the ability for efficient overlap with high loop harmonics through the cross-cut of field lines across many radial oscillations. Under conditions of weak but non-zero dissipation, the energy lost to excitation of torsional resonances equals the work done against the Lorentz force by the spheroidal motions. The singular nature of the interaction acts as an energy sink, giving rise to a global damping of the spheroidal modes that is larger when more resonances are available. The observability of this effect relies on the existence of modes that have large amplitudes both in the core, so that overlaps with torsional resonances are significant, and at the surface, so that the mode can be observed. Evolved stars meet these conditions through a strong coupling of their p- and g-mode cavities, which forms modes of mixed character. 

As already discussed in Section \ref{sec:limits}, the effects of nonlinearity and non-axisymmetry have yet to be dealt with. A number of numerical issues also limit the accuracy of our quantitative results, which would need to be addressed before a detailed exploration of parameter space (e.g.~examining mass and age dependencies) is attempted. Also of interest is to investigate the consequences for our mechanism of modification to the spheroidal mode structure by a strong core field. The recent work of \citet{Lecoanet2016} has presented local calculations of the process, but further work (particularly the formulation of a global description) will be necessary to develop a general theory for the magnetic damping of stellar oscillations that encompasses both strong- and weak-field regimes.

\section*{Acknowledgments}
We thank Mike Proctor and Henrik Latter for helpful comments and discussions. STL is supported by funding from the Cambridge Australia Trust.

%\bibliography{C:/Users/Cleo/Documents/Science/PhD/DAMTP/refs}

\begin{thebibliography}{}

\bibitem[\protect\citeauthoryear{Ballot, Barban \& Van't Veer-Menneret}{Ballot
  et~al.}{2011}]{Ballot2011}
Ballot J.,  Barban C.,    Van't Veer-Menneret C.,  2011, A{\&}A, 531, 124

\bibitem[\protect\citeauthoryear{Braithwaite}{Braithwaite}{2008}]{Braithwaite2008}
Braithwaite J.,  2008, MNRAS, 386, 1947

\bibitem[\protect\citeauthoryear{Braithwaite \& Nordlund}{Braithwaite \&
  Nordlund}{2006}]{Braithwaite2006}
Braithwaite J.,  Nordlund A.,  2006, A\&A, 450, 1077

\bibitem[\protect\citeauthoryear{Braithwaite \& Spruit}{Braithwaite \&
  Spruit}{2004}]{Braithwaite2004}
Braithwaite J.,  Spruit H.~C.,  2004, Nature, 431, 819

\bibitem[\protect\citeauthoryear{Braithwaite \& Spruit}{Braithwaite \&
  Spruit}{2015}]{Braithwaite2015}
Braithwaite J.,  Spruit H.~C.,  2015, Living Reviews, pp 1--57

\bibitem[\protect\citeauthoryear{Brun, Browning \& Toomre}{Brun
  et~al.}{2005}]{Brun2005}
Brun A.~S.,  Browning M.~K.,    Toomre J.,  2005, ApJ, 629, 461

\bibitem[\protect\citeauthoryear{Campbell \& Papaloizou}{Campbell \&
  Papaloizou}{1986}]{Campbell1986}
Campbell C.~G.,  Papaloizou J. C.~B.,  1986, MNRAS, 220, 577

\bibitem[\protect\citeauthoryear{Cantiello, Fuller \& Bildsten}{Cantiello
  et~al.}{2016}]{Cantiello2016}
Cantiello M.,  Fuller J.,    Bildsten L.,  2016, ApJ, 824, 14

\bibitem[\protect\citeauthoryear{Charbonneau \& MacGregor}{Charbonneau \&
  MacGregor}{2001}]{Charbonneau2001}
Charbonneau P.,  MacGregor K.~B.,  2001, ApJ, 559, 1094

\bibitem[\protect\citeauthoryear{Cunha \& Gough}{Cunha \&
  Gough}{2000}]{Cunha2000}
Cunha M.~S.,  Gough D.,  2000, MNRAS, 319, 1020

\bibitem[\protect\citeauthoryear{Deubner \& Gough}{Deubner \&
  Gough}{1984}]{Deubner1984}
Deubner F.-L.,  Gough D.,  1984, ARA{\&}A, 22, 593

\bibitem[\protect\citeauthoryear{Duez, Braithwaite \& Mathis}{Duez
  et~al.}{2010}]{Duez2010}
Duez V.,  Braithwaite J.,    Mathis S.,  2010, ApJ, 724, L34

\bibitem[\protect\citeauthoryear{Duez \& Mathis}{Duez \&
  Mathis}{2010}]{Duez2010a}
Duez V.,  Mathis S.,  2010, A\&A, 517, A58

\bibitem[\protect\citeauthoryear{Dupret et~al.,}{Dupret
  et~al.}{2009}]{Dupret2009_mn2e}
Dupret M.~A.,  et~al., 2009, A{\&}A, 506, 57

\bibitem[\protect\citeauthoryear{Eckart}{Eckart}{1960}]{Eckart1960}
Eckart C.,  1960, Hydrodynamics of oceans and atmospheres.
Pergamon Press

\bibitem[\protect\citeauthoryear{Featherstone, Browning, Brun \&
  Toomre}{Featherstone et~al.}{2009}]{Featherstone2009}
Featherstone N.~A.,  Browning M.~K.,  Brun A.~S.,    Toomre J.,  2009, ApJ,
  705, 1000

\bibitem[\protect\citeauthoryear{Flowers \& Ruderman}{Flowers \&
  Ruderman}{1977}]{Flowers1977}
Flowers E.,  Ruderman M.,  1977, ApJ, 215, 302

\bibitem[\protect\citeauthoryear{Fuller, Cantiello, Stello, Garc\'{i}a \&
  Bildsten}{Fuller et~al.}{2015}]{Fuller2015}
Fuller J.,  Cantiello M.,  Stello D.,  Garc\'{i}a R.~A.,    Bildsten L.,  2015,
  Science, 350, 423

\bibitem[\protect\citeauthoryear{Garc{\'{i}}a et~al.,}{Garc{\'{i}}a
  et~al.}{2014}]{Garcia2014_mn2e}
Garc{\'{i}}a R.~A.,  et~al., 2014, A\&A, 563, A84

\bibitem[\protect\citeauthoryear{Gough}{Gough}{1986}]{Gough1986}
Gough D.~O.,  1986, in Osaki Y.,  ed., Hydrodynamic and Magnetohydrodynamic
  Problems in the Sun and Stars {EBK quantization of stellar waves}.
University of Tokyo, Tokyo, pp 117--143

\bibitem[\protect\citeauthoryear{Houdek \& Dupret}{Houdek \&
  Dupret}{2015}]{Houdek2015}
Houdek G.,  Dupret M.-A.,  2015, Living Reviews in Solar Physics, 12, 1

\bibitem[\protect\citeauthoryear{Lecoanet, Vasil, Fuller, Cantiello \&
  Burns}{Lecoanet et~al.}{2016}]{Lecoanet2016}
Lecoanet D.,  Vasil G.~M.,  Fuller J.,  Cantiello M.,    Burns K.~J.,  2016,
  MNRAS Advance Access

\bibitem[\protect\citeauthoryear{Lee}{Lee}{2007}]{Lee2007}
Lee U.,  2007, MNRAS, 374, 1015

\bibitem[\protect\citeauthoryear{Markey \& Tayler}{Markey \&
  Tayler}{1973}]{Markey1973}
Markey P.,  Tayler R.~J.,  1973, MNRAS, 163, 77

\bibitem[\protect\citeauthoryear{Mestel}{Mestel}{2012}]{Mestel2012}
Mestel L.,  2012, Stellar magnetism, 2nd edn.
Oxford Science Publications

\bibitem[\protect\citeauthoryear{Morel}{Morel}{1997}]{Morel1997}
Morel P.,  1997, A{\&}ASS, 124, 597

\bibitem[\protect\citeauthoryear{Mosser et~al.,}{Mosser
  et~al.}{2011}]{Mosser2011_mn2e}
Mosser B.,  et~al., 2011, A\&A, 525, L9

\bibitem[\protect\citeauthoryear{Mosser et~al.,}{Mosser
  et~al.}{2012a}]{Mosser2012_mn2e}
Mosser B.,  et~al., 2012a, A\&A, 537, A30

\bibitem[\protect\citeauthoryear{Mosser et~al.,}{Mosser
  et~al.}{2012b}]{Mosser2012b_mn2e}
Mosser B.,  et~al., 2012b, A\&A, 548, A10

\bibitem[\protect\citeauthoryear{Mosser et~al.,}{Mosser
  et~al.}{2016}]{Mosser2016}
Mosser B.,  et~al., 2016, A{\&}A, accepted

\bibitem[\protect\citeauthoryear{Osaki}{Osaki}{1975}]{Osaki1975}
Osaki Y.,  1975, PASJ, 27, 237

\bibitem[\protect\citeauthoryear{Prendergast}{Prendergast}{1956}]{Prendergast1956}
Prendergast K.~H.,  1956, ApJ, 123, 498

\bibitem[\protect\citeauthoryear{Proctor \& Gilbert}{Proctor \&
  Gilbert}{1994}]{Proctor1994}
Proctor M.~R.~E.,  Gilbert A.~D.,  1994, Lectures on solar and planetary
  dynamos.
Cambridge University Press

\bibitem[\protect\citeauthoryear{Reese, Rincon \& Rieutord}{Reese
  et~al.}{2004}]{Reese2004}
Reese D.,  Rincon F.,    Rieutord M.,  2004, A\&A, 427, 279

\bibitem[\protect\citeauthoryear{Rincon \& Rieutord}{Rincon \&
  Rieutord}{2003}]{Rincon2003}
Rincon F.,  Rieutord M.,  2003, A\&A, 398, 663

\bibitem[\protect\citeauthoryear{Roberts}{Roberts}{1967}]{Roberts1967}
Roberts P.~H.,  1967, An introduction to magnetohydrodynamics.
Longmans, London

\bibitem[\protect\citeauthoryear{Scuflaire}{Scuflaire}{1974}]{Scuflaire1974}
Scuflaire R.,  1974, A{\&}A, 36, 107

\bibitem[\protect\citeauthoryear{Spitzer}{Spitzer}{1962}]{Spitzer1962}
Spitzer L.,  1962, Physics of fully ionized gases.
John Wiley \& Sons, New York

\bibitem[\protect\citeauthoryear{Spruit \& Bogdan}{Spruit \& Bogdan}{1992}]
{Spruit1992}
Spruit, H.~C.,  Bogdan, T.~J.,  1992, ApJ, 391, L109

\bibitem[\protect\citeauthoryear{Stello, Cantiello, Garc\'{i}a \& Huber}{Stello
  et~al.}{2016}]{Stello2016a}
Stello D.,  Cantiello M.,  Garc\'{i}a R.~A.,    Huber D.,  2016, PASA, 33, e011

\bibitem[\protect\citeauthoryear{Stello et~al.,}{Stello
  et~al.}{2016}]{Stello2016_mn2e}
Stello D.,  et~al., 2016, Nature, 529, 364

\bibitem[\protect\citeauthoryear{Tassoul}{Tassoul}{1980}]{Tassoul1980}
Tassoul M.,  1980, ApJSS, 43, 469

\bibitem[\protect\citeauthoryear{Tayler}{Tayler}{1973}]{Tayler1973}
Tayler R.~J.,  1973, MNRAS, 161, 365

\bibitem[\protect\citeauthoryear{Unno, Osaki, Ando, Saio \& Shibahashi}{Unno
  et~al.}{1989}]{Unno1989}
Unno W.,  Osaki Y.,  Ando H.,  Saio H.,    Shibahashi H.,  1989, Nonradial
  oscillations of stars, 2nd edn.
University of Tokyo Press

\end{thebibliography}
%\bibliographystyle{mn2e}

\end{document}